\documentclass[onecolumn,aps,prd,groupedaddress,nofootinbib,showpacs]{revtex4}
\usepackage{graphicx}
\usepackage{subfigure}
\usepackage{multirow}
\usepackage{amsmath}
\usepackage{bm}
\usepackage{slashed}
\usepackage{epsfig}
\usepackage{amsfonts}
\usepackage{color}
\usepackage{epstopdf}
\usepackage{enumitem}
\usepackage{autobreak}
\allowdisplaybreaks
\newcommand{\ud}{\mathrm{d}}

\newcommand{\jpsi}{{J/\psi}}

\newcommand{\state}[4]{{^{#1}\hspace{-0.6mm}#2_{#3}^{[#4]}}}

\newcommand\CScSa{\state{3}{S}{1}{1}}

\newcommand\mo{{\mathcal O}}

\newcommand\mpp{{\mathcal P}}
\newcommand{\LDMEnpp}[1]{\langle\mpp^{#1}_n\rangle}
\newcommand\mphn{\LDMEnpp{H}}

\newcommand{\LDME}[2]{\langle\mo^{#1}(#2)\rangle}

\newcommand{\LDMEn}[1]{\langle\mo^{#1}_n\rangle}

\newcommand\mohn{\LDMEn{H}}

\def\li{\mathrm{Li}}
\def\co{{\cal O}}

\begin{document}

\title{Analytical calculation for the gluon fragmentation into spin-triplet S-wave quarkonium}% Force line breaks with \\

%\author{}%
%\affiliation{
%}%

\author{Peng Zhang$^{1}$}
\author{Yan-Qing Ma$^{1,2,3}$}
\author{Qian Chen$^{1}$}
\author{Kuang-Ta Chao$^{1,2,3}$}
\affiliation{
$^{1}$School of Physics and State Key Laboratory of Nuclear Physics and
Technology, Peking University, Beijing 100871, China\\
$^{2}$Center for High Energy Physics,
Peking University, Beijing 100871, China\\
$^{3}$Collaborative Innovation Center of Quantum Matter,
Beijing 100871, China
}%
\date{\today}

\begin{abstract}
Fragmentation is the dominant mechanism for hadron production with high transverse momentum. For spin-triplet S-wave heavy quarkonium production, contribution of gluon fragmenting to color-singlet channel has been numerically calculated since 1993. However, there is still no analytic expression available up to now because of its complexity. In this paper, we calculate both polarization-summed and polarized fragmentation functions of gluon fragmenting to a heavy quark-antiquark pair with quantum number $\CScSa$. Our calculations are performed in two different frameworks. One is the widely used nonrelativistic QCD factorization, and the other is the newly proposed soft gluon factorization. In either case, we calculate at both leading order and next-to-leading order in velocity expansion. All of our final results are presented in terms of compact analytic expressions.
\end{abstract}

\pacs{12.38.Bx, 12.39.St, 13.87.Fh, 14.40.Pq}
%\pacs{12.39.St, 13.87.Fh,  14.40.Pq}
%12.38.Bx 	Perturbative calculations
%12.39.St 	Factorization
%13.87.Fh 	Fragmentation into hadrons
%14.40.Pq 	Heavy quarkonia
%11.80.La 	Multiple scattering

%\keywords{Suggested keywords}%Use showqkeys class option if keyword
\maketitle

\section{Introduction}

As heavy quark mass $m_Q$ is much larger than the QCD nonperturbative scale $\Lambda _{QCD}$, the production of heavy quark-antiquark ($Q\bar{Q}$) pair is perturbatively calculable. Due to the binding energy of $Q\bar{Q}$ for a heavy quarkonium being at the order of $\Lambda _{QCD}$,  hadronization of $Q\bar{Q}$ to heavy quarkonium is nonperturbative. Therefore, study of quarkonium production can help to understand both perturbative and nonperturbative physics in QCD. Nevertheless, more than 40 years after the discovery of the $J/\psi$, the production mechanism of heavy quarkonium, the simplest system under strong interaction, is still not well understood.

Recently, two of the present authors proposed a soft gluon factorization (SGF) theory to describe quarkonium production and decay  \cite{Ma:2017xno,machao}. On the one hand, SGF is as rigorous as the currently widely used nonrelativistic QCD factorization (NRQCD) \cite{Bodwin:1994jh}, which means either both of them are correct to all orders in perturbation theory, or both of them are broken down at a sufficient large order in $\alpha_s$ expansion. On the other hand, it was argued that the convergence of velocity expansion in SGF should be much better than that in NRQCD \cite{Ma:2017xno}. Thus, SGF may resolve some difficulties encountered in NRQCD for quarkonium production. In this paper, we use SGF and NRQCD to compute the gluon fragmentation function (FF) to  $J^{PC}=1^{--}$ quarkonia, which is useful for understanding the production of these quarkonia at high transverse momentum $p_T$.
According to QCD collinear factorization \cite{Collins:1989gx}, the inclusive production cross section of a specific hadron $H$ at very high $p_T$ is dominated by the fragmentation mechanism at leading power (LP) \footnote{When $p_T$ is not large enough, next-to-leading power (NLP) contribution will be also important. For quarkonium production, the NLP contribution can also be factorized in terms of perturbative hard part convoluting with double parton fragmentation function \cite{Kang:2011mg,Fleming:2012wy,Kang:2014tta}, which will not be discussed in this paper.},
\begin{equation}
\ud \sigma _{A+B\rightarrow H(p_T)+X} = \sum_i \ud \hat{\sigma} _{A+B \rightarrow i(p_T /z)+X} \otimes D_{i\to H}(z,\mu) + \co (1/p_T^2) \, ,
\end{equation}
where $i$ sums over all quarks and gluons, and $z$ is the light-cone momentum fraction carried by $H$ with respect to the parent parton.
The hard part $d \hat{\sigma} _{A+B \rightarrow i(p_T /z)+X}$ can be calculated perturbatively, while the FF $D_{i\to H}(z,\mu)$, describing the probability distribution of the hadronization from $i$ to $H$, is nonperturbative and universal. The dependence of FF on factorization scale $\mu$ is controlled by the DGLAP evolution equation\cite{Gribov:1972ri,Altarelli:1977zs,Dokshitzer:1977sg},
\begin{equation}
\mu \frac{\ud}{\ud \mu} D_{i\to H}(z,\mu) = \sum_j \int_z^1 \frac{\ud \xi}{\xi} P_{ij} \left( \frac{z}{\xi},\alpha_s(\mu) \right) D_{j\to H}(\xi,\mu) \, ,
\end{equation}
where $P_{ij}$ are splitting functions that can be calculated perturbatively. Based on this evolution, FF at an arbitrary perturbative scale $\mu$ can be determined by FF at an initial scale $\mu_0$.

In the case that $H$ is a heavy quarkonium, there is an intrinsic hard scale $m_Q$ in FFs. Usually, we can choose the initial scale $\mu_0 \gtrsim 2 m_Q $, so that $\ln (\mu_0^2/m_Q^2)-$type logarithms are not large. As $m_Q\gg\Lambda_{\text{QCD}}$, FFs evaluated at $\mu_0$ can be further factorized as perturbative calculable short-distance coefficients (SDCs) multiplied by nonpertubative part at the scale $m_Q v$ and below. If one uses either SGF or NRQCD to do this factorization, one needs to sum over states of intermediate $Q\bar{Q}$ pair, which are usually expressed as spectroscopic notation $\state{{2S+1}}{L}{J}{c}$ with $c=1~\text{or}~8$ to  denote color singlet or color octet. For the gluon fragmentation to $1^{--}$ quarkonium, like the $\jpsi$, $\psi(2S)$ and $\Upsilon(nS)$, the dominant contribution comes from $\state{3}{S}{1}{1}$ intermediate state according to the velocity scaling rule \cite{Bodwin:1994jh}.

In the SGF framework, calculation of fragmentation function from gluon to $\state{3}{S}{1}{1}$ intermediate state is still absent. In the NRQCD framework, SDCs of gluon fragmenting into polarization summed $\state{3}{S}{1}{1}$ intermediate state has been calculated numerically in Refs.~\cite{Braaten:1993rw,Braaten:1995cj} for $v^0$ contribution and in Ref.~\cite{Bodwin:2003wh} for $v^2$ correction. However, analytical results are still absent because of the complexity of the problem.
One reason is that there are two gluons emitted in the final state, so the phase space integral is similar to two-loops integral. The other reason is that light-cone momentum is involved in the definition of fragmentation function, which makes the phase space integral more complicated than usual. In Refs. \cite{Braaten:1993rw,Braaten:1995cj}, there is a four-dimensional integral left for numerical computing. In Ref. \cite{Bodwin:2003wh}, the authors make some transformation of the variables and analytically integrate out two more dimensions, but there is still a two-dimensional integral that has to be calculated numerically.
Besides, there is no calculation of polarized SDCs based on the definition of fragmentation function, of which the result is useful for understanding the polarization puzzle \cite{Abulencia:2007us,Butenschoen:2012px,Chao:2012iv,Gong:2012ug,Bodwin:2014gia}.

In this paper, we analytically calculate SDCs of gluon fragmenting into $\state{3}{S}{1}{1}$ state separately in SGF and NRQCD frameworks. We include both  $v^0$ contribution and  $v^2$ contribution in nonrelativistic expansion. In all cases, we provide transversely polarized SDCs in addition to polarization summed SDCs, while longitudinally polarized SDCs can be obtained by their difference.

The rest of the paper is organized as follows. In Sec.~\ref{sec:def}, we first introduce the definition of gluon FF, and then describe how to apply SGF and NRQCD to calculate the FF in detail. The resulting expressions are complicated phase space integrals. In Sec.~\ref{sec:cal}, we use integration-by-part (IBP) method \cite{Chetyrkin:1981qh,Smirnov:2012gma,Smirnov:2014hma,Lee:2013mka} to express these phase space integrals in terms of some bases, which are called master integrals. We then calculate these master integrals. Almost all master integrals can be easily calculated except one, which we calculate by constructing and solving a differential equation with a trivial initial condition. Then we exhibit the analytical results and the large $z$ behaviour of the FFs. Finally, we present numerical results and a discussion in Sec.~\ref{sec:summary}. Some coefficients of analytical results calculated in this paper are given in the Appendix.

\section{Factorization of Quarkonium Fragmention Functions}\label{sec:def}

\subsection{Definition of fragmentation functions}

In this paper, we use light-cone coordinates where a four-vector $V$ can be expressed as
\begin{align}
\begin{split}
	V   & = (V^+,V^-,\boldsymbol{V}_{\perp})=(V^+,V^-,V^1,V^2) \, ,  \\
	V^+ & = (V^0+V^3)/\sqrt{2} \, , \\
	V^- & = (V^0-V^3)/\sqrt{2} \, .
\end{split}
\end{align}
The scalar product of two four-vector $V$ and $W$ then becomes
\begin{equation}
V \cdot W = V^+ W^- + V^- W^+ - \boldsymbol{V}_{\perp} \cdot \boldsymbol{W}_{\perp} \, .
\end{equation}
We introduce a light-like vector $n=(0,1,\boldsymbol{0}_{\perp})$, so that $n\cdot V=V^+$.

The Collins-Soper definition of FF for a gluon fragmenting into a hadron (quarkonium) is given by \cite{Collins:1981uw}
\begin{align}\label{eq:defFF}
\begin{split}
    D_{g \rightarrow H}(z,\mu_0)=
    & \frac{-g_{\mu\nu}z^{D-3}}{2 \pi P_c^{+}(N_{c}^{2}-1)(D-2)} \int_{-\infty}^{+\infty}\mathrm{d}x^{-} e^{-i z P_c^{+} x^{-}} \\
    & \times \langle 0 | G_{c}^{+\mu}(0) \mathcal{E}^{\dag}(0,0,\boldsymbol{0}_{\perp})_{cb} \mathcal{P}_{H(P_H)} \mathcal{E}(0,x^{-},\boldsymbol{0}_{\perp})_{ba} G_{a}^{+\nu}(0,x^{-},\boldsymbol{0}_{\perp}) | 0 \rangle \, ,
\end{split}
\end{align}
where $G^{\mu\nu}$ is the gluon field-strength operator, $P_H$ and $P_c$ are respectively the momenta of the fragmenting hadron and initial virtual gluon, and $z$ is the ``$+$'' momentum fraction of the initial virtual gluon carried by the hadron. It is convenient to choose the frame so that the hadron has zero transverse momentum, $P_H = (z P_c^+, M_H^2/(2 z P_c^+),\boldsymbol{0}_{\perp})$, where $M_H$ is the mass of the hadron.
The projection operator $\mathcal{P}_{H(P_H)}$ is given by
\begin{equation}\label{eq:projectH}
\mathcal{P}_{H(P_H)} = \sum_X |H(P_H)+X \rangle \langle H(P_H)+X|\,,
\end{equation}
where $X$ sums over all unobserved particles.
The gauge link $\mathcal{E}(x^{-})$ is an eikonal operator that involves a path-ordered exponential of gluon field operators along a light-like path,
\begin{equation}
  \mathcal{E}(0,x^{-},\boldsymbol{0}_{\perp})_{ba}= \mathrm{P} \, \text{exp} \left[+i g_s \int_{x^{-}}^{\infty}\mathrm{d}z^{-} A^{+}(0,z^{-},\boldsymbol{0}_{\perp}) \right]_{ba} \, ,
\end{equation}
where $g_s=\sqrt{4 \pi \alpha_s}$ is the QCD coupling constant and $A^{\mu}(x)$ is the matrix-valued gluon field in the adjoint representation: $[A^{\mu}(x)]_{ac} = i f^{abc} A^{\mu}_{b}(x)$. In the light-cone gauge $A^+=A\cdot n=0$, the gauge link $\mathcal{E}(0,x^{-},\boldsymbol{0}_{\perp})$ becomes 1 and thus it does not show up in the Feynman diagrams. In fact, for the problem studied in this paper, the gauge link has no contribution even if we work in Feynman gauge.

\subsection{Applying SGF to fragmentation functions}

The one-dimensional SGF for gluon fragmenting to quarkonium $H$ is given by \cite{Ma:2017xno}
\begin{equation}\label{eq:SGF}
D_{g\to H}(z,\mu_0) = \sum_n\int dr\, d_{n} (z/r,M_H/r,m_Q,\mu_0)\,r  F_n^H(r) \, ,
\end{equation}
where $n=\state{{2S+1}}{L}{J}{c}$ is in spectroscopic notation to denote quantum numbers of the intermediate $Q\bar{Q}$ pair, $d_{n} (z/r,M_H/r,m_Q,\mu_0)$ are SDCs to produce a $Q\bar{Q}$ pair with invariant mass $M_H/r$ and quantum number $n$,   $F_n^H(r)$ are one-dimensional soft gluon distributions (SGDs) defined by four-dimensional SGDs
\begin{align}
F_{n}^H(r)=\int \frac{d^4P}{(2\pi)^4} \delta(r-\sqrt{P_H^2/P^2})\,  F^H_{n}(P,P_H),
\end{align}
and four-dimensional SGDs are defined as expectation values of bilocal operators in QCD vacuum,
\begin{align}\label{eq:SGDs}
\begin{split}
&F^H_{n}(P,P_H)=\int d^4x\, e^{iP\cdot x}\, \langle 0| \bar\psi(0)\Gamma_{n}^\prime \mathcal{E}^\dagger(0)\psi(0) \mathcal{P}_{H(P_H)} \bar\psi(x) \Gamma_{n}\mathcal{E}(x)\psi(x) |0\rangle,
\end{split}
\end{align}
where $\Gamma$ and $\Gamma^\prime$ are color and angular momentum projection operators that define $n$, $\mathcal{E}(x)$ are gauge links that enable gauge invariance  \cite{machao,Ma:2017xno}.

\begin{figure}[htb!]
 \begin{center}
 \includegraphics[width=0.4\textwidth]{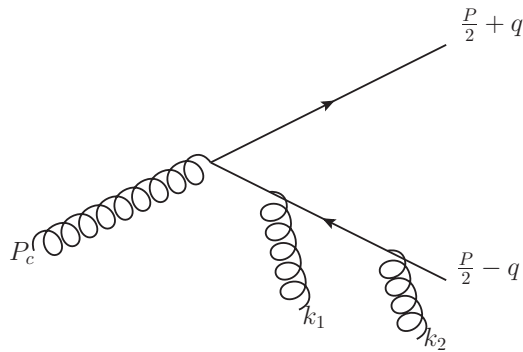}
  \caption{One typical diagram for the gluon fragmenting into  $\state{3}{S}{1}{1} \  Q\bar{Q}$ pair in the light-cone gauge at LO order in $\alpha_s$. The other diagrams are obtained by permutation. \label{fig:FeynmanDiagram}}
 \end{center}
\end{figure}

As mentioned in the introduction, for gluon fragmenting to $1^{--}$ quarkonium in this paper we keep only $n=\state{3}{S}{1}{1}$ intermediate $Q\bar{Q}$ state, we thus suppress the subscript $n$ in the rest of this paper. Then the lowest order in $\alpha_s$ expansion of $d (z/r,M_H/r,m_Q,\mu_0)$ is described by Feynman diagrams of a virtual gluon decaying to a $Q\bar{Q}$ pair with quantum number $\state{3}{S}{1}{1}$ combined with two more gluons, as shown in Fig.~\ref{fig:FeynmanDiagram} , which can be formally defined as the lowest order in $\alpha_s$ expansion of the following matrix element,
\begin{align}\label{eq:defsdc}
\begin{split}
    & d^{\infty}(z,M,m_Q,\mu_0)= \frac{-g_{\mu\nu}z^{D-3}}{2 \pi P_c^{+}(N_{c}^{2}-1)(D-2)} \int_{-\infty}^{+\infty}\mathrm{d}x^{-} e^{-i z P_c^{+} x^{-}} \\
    & \times \langle 0 | G_{c}^{+\mu}(0) \mathcal{E}^{\dag}(0,0,\boldsymbol{0}_{\perp})_{cb} |Q\bar{Q}(\state{3}{S}{1}{1})+g+g \rangle \langle Q\bar{Q}(\state{3}{S}{1}{1})+g+g| \mathcal{E}(0,x^{-},\boldsymbol{0}_{\perp})_{ba} G_{a}^{+\nu}(0,x^{-},\boldsymbol{0}_{\perp}) | 0 \rangle \, ,
\end{split}
\end{align}
where $M$ is the invariant mass of the $Q\bar{Q}$ pair.

In the momentum space, we have the lowest order in $\alpha_s$ expansion
\begin{equation}\label{eq:sdcdef0}
	d(z,M,m_Q,\mu_0) = \frac{N_{\mathrm{CS}}}{D-1} \int \mathrm{d} \Phi \, \left| \mathcal{M}(P, k_i, m_Q) \right|^2 \, ,
\end{equation}
where $N_{\mathrm{CS}}=\frac{z^{D-2}}{(N_{c}^{2}-1)(D-2)} $, $D=4$ is the space-time dimension, $P$ is the total momentum of the $Q\bar{Q}$ pair, $k_i$ ($i=1,2$) is the momentum of the $i$th final-state gluon, and final-state phase space is defined as
\begin{align}\label{eq:phase}
\begin{split}
  \mathrm{d} \Phi &=
  	\frac {1}{2!}
  	\delta \left( z - \frac{P^{+}}{P_{c}^{+}} \right)
    (2\pi)^{D} \delta^{D} \left(
    P_{c} - P - k_1-k_2 \right)
    \frac{\mathrm{d}^{D} P_{c}}{(2\pi)^{D}}
    \prod_{i=1}^{2} \frac{\mathrm{d} k_{i}^{+}}{4\pi k_{i}^{+}}
    \frac{\mathrm{d}^{D-2} k_{i\perp}}{(2\pi)^{D-2}}
    \theta(k_{i}^{+}) \\
    &=
	\frac {P^{+}}{z^{2} 2!}
      \delta \left( \frac{1-z}{z} P^{+} - k_{1}^{+} - k_{2}^{+}\right)
      \prod_{i=1}^{2} \frac{\mathrm{d} k_{i}^{+}}{4\pi k_{i}^{+}}
      \frac{\mathrm{d}^{D-2} k_{i\perp}}{(2\pi)^{D-2}}
      \theta(k_{i}^{+}) \,
\end{split}
\end{align}
where $2!$ is the symmetry factor for identical gluons in the final state.
The matrix elements $\mathcal{M}({P},{k}_{i},m_Q)$ are defined as
\begin{align}
\left| \mathcal{M}({P},{k}_{i},m_Q) \right|^2=\sum_{\lambda \lambda_1 \lambda_2 \lambda_3} \left| \mathcal{M}_{\lambda \lambda_1 \lambda_2 \lambda_3}(P,k_{i},m_Q) \right|^2,
\end{align}
where  $\lambda$ and $\lambda_i$ ($i=1, 2, 3$) are polarizations of the $Q\bar{Q}$ pair and gluons, respectively, and $\mathcal{M}_{\lambda \lambda_1 \lambda_2 \lambda_3}(P,k_{i},m_Q)$ is the amplitude to produce a $Q\bar{Q}$ pair with momentum $P$ and quantum numbers  $\state{3}{S}{1}{1}$. Summation over all polarizations of the heavy quark pair with momentum $P$ gives
\begin{align}\label{eq:polarsum}
	I^{\alpha\beta}(P)=\sum_{\lambda=0,\pm 1} \epsilon_\lambda^\alpha& \epsilon_\lambda^{*\beta} = -g^{\alpha \beta}
			+ \frac{P^{\alpha} P^{\beta}} {P^{2}} \, ,		
\end{align}
and summation over all polarizations of gluons or summation over transverse polarizations of the heavy quark pair with momentum $k$ gives
\begin{align}\label{eq:polarsumT}
	\sum_{\lambda_i=\pm 1} \epsilon_{\lambda_i}^\mu& \epsilon_{\lambda_i}^{*\nu} = -g^{\mu \nu}
			+ \frac{k^{\mu} n^{\nu} + k^{\nu} n^{\mu}} {k^{+}}
			- \frac{k^2 n^{\mu} n^{\nu}} {(k^{+})^{2}} \, .
\end{align}
In the above, polarizations are quantized along the vector $n$.

The amplitude to produce a $\state{3}{S}{1}{1}$ state can be obtained by
\begin{align}\label{eq:amplitude}
\mathcal{M}_{\lambda \lambda_1 \lambda_2 \lambda_3}(P,k_{i},m_Q)=\int \ud^2 \Omega\, \text{Tr}\left[ \Gamma_\lambda \mathcal{M}_{\lambda_1 \lambda_2 \lambda_3}(P,k_{i},q,m_Q)  \right]\, ,
\end{align}
where  $\mathcal{M}_{\lambda_1 \lambda_2 \lambda_3}(P,k_{i},q,m_Q)$ is the amplitude to produce an open $Q\bar{Q}$ pair, with momenta $p=P/2+q$ for $Q$ and $\overline{p}=P/2-q$ for $\bar Q$, and $\Gamma_\lambda $ is used to project the $Q\bar{Q}$ pair to $\state{3}{S}{1}{1}$ state, with definition
\begin{equation}
	\Gamma_{\lambda} = \frac{1} {\sqrt{N_{c}}}
	    \frac{1} {\sqrt{2 E} (E + m_Q)}
		(\slashed{\overline{p}} - m_Q)
		\frac{2 E - \slashed{P}}{4 E}
		\slashed{\epsilon}_\lambda
		\frac{2 E + \slashed{P}}{4 E}
		(\slashed{p} - m_Q)	\, ,
\end{equation}
where $\epsilon_\lambda^\mu$ are polarization vectors. As both $p$ and $\overline{p}$ are approximated to be on mass shell  \cite{machao,Ma:2017xno}, we have
\begin{align}\label{eq:relation}
P\cdot q =0,  \quad \quad M^2=P^2=4E^2=4(m_Q^2-q^2).
\end{align}\
As a result, the four-momentum $q$ has only two degrees of freedom, which are chosen to be the two-dimensional spatial angles $\Omega$ at the rest frame of $P$. After integration over spatial angles, the obtained $\mathcal{M}_{\lambda \lambda_1 \lambda_2 \lambda_3}(P,k_{i},m_Q)$ in Eq.~\eqref{eq:amplitude} has no dependence on $q$ any more.

Note that the dominant contribution of Eq.~\eqref{eq:SGF} comes from the region where $1\gg1-r^2=1-M_H^2/M^2\approx1-4m_Q^2/M^2=-4q^2/M^2$  \cite{machao,Ma:2017xno}, thus we can simplify SDCs by expanding $q^2/M^2$ .
In the SGF framework, this expansion is obtained by first expressing $m_Q^2={M^2/4+q^2}$, and then  fixing $M$ but expanding $q^2$ at the origin  \footnote{It is needed to pointed out that $q$ in the amplitude is not the same as that, say $q^\prime$, in the complex conjugate of the amplitude, but $q^2=q^{\prime2}$.}.
%\begin{align}\label{eq:sdcdef}
%\begin{split}
%d_{g \to Q \overline{Q}(\state{3}{S}{1}{1})}(z,M,m_Q,\mu_0) &= d(z,M,M/2,\mu_0) + q^2 \,\frac{\partial }{\partial m_Q^2}d(z,M,m_Q,\mu_0)\big|_{m_Q=M/2} + \cdots\\
%&\equiv d^{(0)}(z) + \frac{-4{q}^2}{M^2} d^{(2)}(z) +\cdots \, .
%\end{split}
%\end{align}
Because neither phase space integration nor polarization vectors depend on $m_Q$ and $q$, the above expansion can be achieved by a similar expansion of the amplitude in Eq.~\eqref{eq:amplitude},
\begin{equation}\label{eq:amplitudeExp}
	\mathcal{M}_{\lambda \lambda_1 \lambda_2 \lambda_3}(P,k_{i},m_Q) =
		\mathcal{M}_{\lambda \lambda_1 \lambda_2 \lambda_3}^{(0)}(P,k_{i})
		-q^2 \mathcal{M}_{\lambda \lambda_1 \lambda_2 \lambda_3}^{(2)}(P,k_{i})+O(q^4)\, ,
\end{equation}
with
\begin{align}\label{eq:amplitudeExp2}
\begin{split}
	\mathcal{M}_{\lambda \lambda_1 \lambda_2 \lambda_3}^{(0)}(P,k_{i})
	& = \text{Tr}\left[ \Gamma_\lambda \mathcal{M}_{\lambda_1 \lambda_2 \lambda_3}(P,k_{i},0,M/2)  \right] \, ,
	\\
	\mathcal{M}_{\lambda \lambda_1 \lambda_2 \lambda_3}^{(2)}(P,k_{i})
	& = \frac{I^{\mu\nu}(P)}{2(D-1)}
		\left\{\frac{\partial^2 }{\partial q^\mu \partial q^\nu}
		\text{Tr}\left[ \Gamma_\lambda \mathcal{M}_{\lambda_1 \lambda_2 \lambda_3}\left(P,k_{i},q,\sqrt{\frac{M^2}{4}+q^2}\right)
		\right]\Bigg|_{q=0}\right\}\, ,
\end{split}
\end{align}
where $I^{\mu\nu}(P)$ is defined in Eq.~\eqref{eq:polarsum}. With this expansion, SDCs have expansion
\begin{align}\label{eq:sdcExp}
	d(z,M,m_Q,\mu_0) =& \frac{N_{\mathrm{CS}}}{D-1} \int \mathrm{d} \Phi \, \sum_{\lambda \lambda_1 \lambda_2 \lambda_3} \left| \mathcal{M}_{\lambda \lambda_1 \lambda_2 \lambda_3}^{(0)}(P,k_{i}) \right|^2 - q^2 \left[ \mathcal{M}_{\lambda \lambda_1 \lambda_2 \lambda_3}^{(0)}(P,k_{i}) \mathcal{M}_{\lambda \lambda_1 \lambda_2 \lambda_3}^{*(2)}(P,k_{i})+ c.c.\right] + O(q^4) \notag\\
	\equiv& d^{(0)}(z, M,\mu_0 ) - \frac{4 q^2}{M^2} d^{(2)}(z, M ,\mu_0) + O(q^4).
\end{align}

Similarly, if we sum over only transerve polarizations of the $Q\bar{Q}$ pair by using the projection operator in Eq.~\eqref{eq:polarsumT} instead of that in Eq.~\eqref{eq:polarsum}, we can obtain transversely polarized SDCs
\begin{align}\label{eq:sdcTExp}
	d_T(z,M,m_Q,\mu_0) = d_T^{(0)}(z, M,\mu_0 ) - \frac{4 q^2}{M^2} d_T^{(2)}(z, M ,\mu_0) + O(q^4).
\end{align}
Longitudinal polarized SDCs can be obtained by subtracting out transversely polarized SDCs from corresponding polarization-summed SDCs.

\subsection{Applying NRQCD to fragmentation functions}

While if applying the NRQCD, we get
\begin{equation}
D_{g\to H}(z,\mu_0) = \sum_n d_n^O (z,2m_Q,\mu_0) \mohn + d_n^{P} (z,2m_Q,\mu_0) \mphn +\cdots\, ,
\end{equation}
where $d_{n}^{O,P} (z,2m_Q,\mu_0)$ are SDCs to produce a $Q\bar{Q}$ pair with invariant mass $2m_Q$ and quantum numbers $n$, and $\mohn$ and $\mphn$ are respectively NRQCD long-distance matrix elements (LDMEs) at first and second order in $v^2$ expansion \cite{Bodwin:1994jh}, which can be expressed as the vacuum expectation value of a four-fermion operator in NRQCD vacuum
\begin{align}
\mohn &= \langle 0 | \chi ^{\dag} \kappa _n \psi \mathcal{P}_{H(P)} \psi ^{\dag} \kappa'_n \chi |0 \rangle \, ,\\
\mphn &= \langle 0 |\frac{1}{2}\left[ \chi ^{\dag} \kappa _n \psi \mathcal{P}_{H(P)} \psi ^{\dag} \kappa'_n (-\frac{i}{2} \overleftrightarrow{\mathbf{D}})^2 \chi + h.c. \right]|0 \rangle \, ,
\end{align}
where $\psi ^{\dag}$ and $\chi$ are the two-component operators to creat a heavy quark and a heavy antiquark, respectively, and $\kappa _n$ and $\kappa'_n$ are combinations of Pauli and color matrices. These LDMEs are defined in the rest frame of $H$ and expected to be universal. If the hadron $H$ is the free $Q \bar{Q}$ pair, we have $\mphn=(-q^2/m_Q^2) \mohn$. As mentioned above, we only consider $n=\CScSa$ intermediate state, and thus will drop the subscript $n$ in the following.

The calculations of $d^O$ and $d^P$ in NRQCD are very similar to the calculation of $d^{(0)}$ and $d^{(2)}$ in SGF defined in Eq.~\eqref{eq:sdcExp}.
The only difference is that, in the NRQCD, one expands $q^2$ with fixed $m_Q$ but not $M$, which implies that phase space also needs to be expanded. For this purpose, we first extract the dependence on $q$ explicitly by rescaling momenta in the delta function in Eq.~\eqref{eq:phase} by $M$ as following,
\begin{equation} \label{eq:dless}
  \hat{P} = \frac {P}{M} \, , \quad
  \hat{k}_i = \frac {k_i}{M} \, .
\end{equation}
Thus the phase space in Eq.~\eqref{eq:phase} changes to
\begin{equation}
	\ud \Phi = M^{4} \ud \hat{\Phi} \, ,
\end{equation}
where $\ud \hat{\Phi}$ is the same as $\ud \Phi$ except that momenta in it have been changed to the dimensionless ones, and therefore it has no dependence on $q$.
If we further denote
\begin{equation}
	\hat{\mathcal{M}}_{\lambda \lambda_1 \lambda_2 \lambda_3}(\hat{P},\hat{k}_{i},m_Q) = M^2 \mathcal{M}_{\lambda \lambda_1 \lambda_2 \lambda_3}(M \hat{P}, M \hat{k}_{i},m_Q) \, ,
\end{equation}
we get a similar relation as that in Eq.~\eqref{eq:sdcdef0},
\begin{equation}
	d(z,M,m_Q,\mu_0) = \frac{N_{\mathrm{CS}}}{D-1} \int \mathrm{d} \hat{\Phi} \, \left| \hat{\mathcal{M}}(\hat{P},\hat{k}_i, m_Q) \right|^2 \,.
\end{equation}
Then the expansion of amplitude $\hat{\mathcal{M}}_{\lambda \lambda_1 \lambda_2 \lambda_3}(\hat{P},\hat{k}_{i},m_Q)$ can be achieved similarly as that in Eq.~\eqref{eq:amplitudeExp} and Eq.~\eqref{eq:amplitudeExp2}, except that we express $M^2=4(m_Q^2-q^2)$ and fix $m_Q$.
Eventually, we get
\begin{align}\label{eq:NRsdcExp}
\begin{split}
	d(z,M,m_Q,\mu_0) = & \frac{N_{\mathrm{CS}} }{D-1} \int \mathrm{d} \hat\Phi\, \sum_{\lambda \lambda_1 \lambda_2 \lambda_3}
						\left| \mathcal{\hat{M}}_{\lambda \lambda_1 \lambda_2 \lambda_3}^{(0)}(\hat{P},\hat{k}_{i},m_Q) \right|^2 \\
					   & \phantom{\frac{N_{\mathrm{CS}} }{D-1} \int \mathrm{d} \hat\Phi\, \sum_{\lambda \lambda_1 \lambda_2 \lambda_3} }
						- q^2 \left[ \mathcal{\hat{M}}_{\lambda \lambda_1 \lambda_2 \lambda_3}^{(0)}(\hat{P},\hat{k}_{i},m_Q)
						\mathcal{\hat{M}}_{\lambda \lambda_1 \lambda_2 \lambda_3}^{*(2)}(\hat{P},\hat{k}_{i},m_Q)+ c.c.\right] + O(q^4) \\
		\equiv& d^{O}(z, 2m_Q,\mu_0 ) -\frac{q^2}{m_Q^2} d^{P}(z, 2m_Q ,\mu_0) + O(q^4) \, .
\end{split}
\end{align}
Clearly, we have the relation
\begin{align}\label{eq:relationU}
d^{O}(z, M,\mu_0 )=d^{(0)}(z,M,\mu_0),
\end{align}
but $d^{P}(z, M,\mu_0 )$ is different from $d^{(2)}(z,M,\mu_0)$.

Again, we can obtain transversely polarized SDCs
\begin{align}\label{eq:NRsdcTExp}
	d_T(z,M,m_Q,\mu_0) = d_T^{O}(z, 2m_Q,\mu_0 ) -\frac{q^2}{m_Q^2} d_T^{P}(z, 2m_Q ,\mu_0) + O(q^4),
\end{align}
where we also have
\begin{align}\label{eq:relationT}
d_T^{O}(z, M,\mu_0 )=d_T^{(0)}(z,M,\mu_0).
\end{align}

\section{Calculation of the short-distance coefficient}\label{sec:cal}

For the process of gluon fragmenting to spin-triplet color-singlet S-wave quarkonium at LO in $\alpha_s$, there are two soft gluons in the final state as shown in Fig.~\ref{fig:FeynmanDiagram}. We denote the ``$+$'' component of the first gluon as $k_1^+ = (1-z) z_1 P_c^+ $, then for the second gluon we have $k_2^+ = (1-z)(1-z_1) P_c^+ $. To simplify our notation, we will use only dimensionless momenta defined in Eq.~\eqref{eq:dless} but omit the superscript `` $\hat{}$ '' in the rest of this paper.

According to Sec.~\ref{sec:def}, calculation of SDCs can be decomposed into the sum of a series of integrals with the form
\begin{equation}\label{eq:iniff}
 	\int \mathrm{d} \Phi
  	f(z,z_{1})
  	\frac {(k_1 \cdot k_2)^{n_{5}}}{E_1^{n_{1}} E_2^{n_{2}} E_3^{n_{3}} E_4^{n_{4}}} \, ,
\end{equation}
where $n_i>0(i=1,2,3,4,5)$,  $f(z,z_{1})$ is fractional polynomials with respect to $z$ and $z_1$, and
\begin{equation}\label{eq:defdeno1}
  E_1 = k_1 \cdot P \, , \
  E_2 = k_2 \cdot P \, , \
  E_3 = 2 k_1 \cdot k_2 + k_1 \cdot P + k_2 \cdot P \, , \
  E_4 = 1 + 2 k_1 \cdot k_2 + 2 k_1 \cdot P + 2 k_2 \cdot P \, .
\end{equation}
We note that $z_1$ does not appear in the denominators, which is because, as we pointed out, the gauge link in the definition of FFs has no contribution in our case.

\subsection{Reduction to Master Integrals}

Calculating general integrations in Eq.~\eqref{eq:iniff} analytically is not an easy task, and only numerical results are available in literature \cite{Braaten:1993rw,Braaten:1995cj,Bodwin:2003wh}. To perform them analytically, we employ the IBP reduction method \cite{Chetyrkin:1981qh,Smirnov:2012gma,Smirnov:2014hma,Lee:2013mka} that are widely used for high loops calculation. Especially, we will use the program FIRE \cite{Smirnov:2014hma}.
Feynman integrals that can be reduced by FIRE can be generally expressed as
\begin{equation}\label{eq:defhloop}
	F(a_{1}, \ldots ,a_{n}) =
		\idotsint \,
		\frac{\ud^{D} l_{1} \ldots \ud^{D} l_{h}}
			{D_{1}^{a_{1}} \ldots D_{n}^{a_{n}}} \, ,
\end{equation}
where $a_i\,(i=1, \ldots, n)$ are integers that can be either positive or negative, and denominators $D_i \,( i=1, \ldots, n)$ are linear functions with respect to scalar products of loop momenta $l_{i}\, (i=1,\ldots,h)$ and external mementa. The program FIRE, by employing IBP, can reduce these complex integrals into limited number of simpler integrals which are called master integrals.
%We find that the loop momenta are changing within the scope of the infinite.

Nevertheless, integrations in Eq.~\eqref{eq:iniff} are not directly handleable by FIRE because there are delta functions in the phase space, which becomes more clearly if we rewrite the phase space as
\begin{equation}\label{eq:phase3}
	\mathrm{d} \Phi =
        \frac{\mathrm{d}^{D} k_1}{(2\pi)^{D}}
        \frac{\mathrm{d}^{D} k_2}{(2\pi)^{D}}
        \frac{P \cdot n}{z^{2} 2!}
        \delta _+ (k_1^{2})
        \delta _+ (k_2^{2})
        \delta \left(
        	k_1 \cdot n + k_2 \cdot n - \frac{1-z}{z} P \cdot n
        	\right) \, ,
\end{equation}
where subscript ``$+$'' of a delta function means that energy of the momentum inside the delta function is positive. To make delta functions handleable by FIRE, we rewrite a delta function as
\begin{equation}
	\delta (x) =
		\frac{i}{2 \pi}
		\lim_{\varepsilon \rightarrow 0}
		\left(
		\frac{1}{x + i \varepsilon}
		- \frac{1}{x - i \varepsilon}
		\right) \, ,
\end{equation}
which changes the delta function to a propagator denominator. We can further identify $z_1=z \, k_1 \cdot n / (1-z) P \cdot n$, and choose the following notations
\begin{equation}\label{eq:defdeno2}
	E_5 = k_1^2 + i \varepsilon \, , \
	E_6 = k_2^2 + i \varepsilon \, , \
	E_7 = k_1 \cdot n + k_2 \cdot n - \frac{1-z}{z} P \cdot n + i \varepsilon \, , \
	E_8 = k_1 \cdot n \, ,
\end{equation}
then integrals in Eq.~\eqref{eq:iniff} are casted to
\begin{equation}\label{eq:intfire}
	\int
	\frac{\mathrm{d}^{D} k_1}{(2\pi)^{D}}
    \frac{\mathrm{d}^{D} k_2}{(2\pi)^{D}}
    f(z)
    \frac{P \cdot n}{z^{2} 2!}
    \left ( \frac{z}{(1-z) P \cdot n} \right)^{n_{6}}
    \left ( \frac{i}{2 \pi} \right )^{3}
    \frac{(k_1 \cdot k_2)^{n_{5}} E_8^{n_{6}}}{E_1^{n_{1}} E_2^{n_{2}} E_3^{n_{3}} E_4^{n_{4}} E_5 E_6 E_7 } \, ,
\end{equation}
together with 7 other kinds of integrals with similar form except that some of small imaginary parts of $E_5,\, E_6,\, E_7$ change from `` $+i\varepsilon$ '' to `` $-i\varepsilon$ ''. Since IBP reduction is independent of the small imaginary part, these 8 kinds of integrals have similar reduced results. Therefore, after reduction, we can change $E_5,\, E_6,\, E_7$ back to corresponding delta functions, and thus we can obtain master integrals of Eq.~\eqref{eq:iniff}. One important point is that any master integral with non-positive power of $E_5,\, E_6,\, E_7$ must be canceled by other master integrals reduced by the other 7 kinds of integrals. Combining with the fact that powers of $E_5,\, E_6,\, E_7$ can be always chosen to no larger than 1, our obtained master integrals have the same phase space integration as that in Eq.~\eqref{eq:iniff}.

The denominators $E_1, \cdots, E_7$  in Eq.~\eqref{eq:intfire} are linearly dependent, which can be easily changed to be linearly independent with the same integrations structure. We further add a denominator $E_8$ to some integrals to make them complete. After reduction by applying FIRE \cite{Smirnov:2014hma}, SDC, say $d^{(0)}$, becomes
\begin{equation} \label{eq:sdcmi}
	d^{(0)} (z,M,\mu_0) = \sum_{a=1}^{13} f_{a} (z, \epsilon)I_{a} \, ,
\end{equation}
where coefficients $f_{a}$ are fractional polynomials in terms of  $z$, which can be expanded in powers of $\epsilon$, and master integrals $I_{a}$ can be defined as
\begin{equation}\label{eq:defmi}
	I_{a} = \int \ud \Phi \, F_{a}
		  = \frac{1}{(4 \pi)^2 z (1 -z ) 2!} \int_{0}^{1} \frac{\mathrm{d} z_{1}}{z_{1} (1 - z_{1})}
    		\iint \frac{\mathrm{d}^{D-2} k_{1\perp}}{(2\pi)^{D-2}} \frac{\mathrm{d}^{D-2} k_{2\perp}}{(2\pi)^{D-2}}
    		F_{a} \, ,
\end{equation}
with $F_{a} (a=1,\ldots,13) $ choosing from
\begin{equation}\label{eq:allmi}
	\frac{1}{E_3} \, , \
	\frac{1}{E_4} \, , \
	\frac{1}{E_1 E_2} \, , \
	\frac{1}{E_1 E_3} \, , \
	\frac{1}{E_1 E_4} \, , \
	\frac{E_2}{E_1 E_3} \, , \
	\frac{E_4}{E_1 E_3} \, , \
	\frac{E_2}{E_1 E_4} \, , \
	\frac{E_3}{E_1 E_4} \, , \
	\frac{1}{E_1 E_3^{2}} \, , \
	\frac{1}{E_1^{2} E_4} \, , \
	\frac{1}{E_3 E_4} \, , \
	\frac{1}{E_1 E_2 E_4} \, ,
\end{equation}
where $E_i (i=1,\ldots,4)$ are defined in Eq.~\eqref{eq:defdeno1}.

\subsection{Calculation of Master Integrals}

Calculation of SDCs is now reduced to calculation of the thirteen master integrals defined in Eq.~\eqref{eq:defmi}.
Among them, each of the first 11 master integrals involves only one denominator that has cross term $k_1\cdot k_2$. In this case, the cross term can be removed by shifting $k_2$, and then we can integrate over $k_{2\perp}$, $k_{1\perp}$ and $z_1$ sequentially. For the 12th master integral, as both of its denominators depend on $k_1\cdot k_2$, we can first do a Feynman parametrization, and then integrate over $k_{2\perp}$, $k_{1\perp}$, Feynman parameter, and $z_1$ sequentially. Although they are easy to calculate, expressions of the first 12 master integrals are quite long, we will not list them in this paper. The most complicated master integral is the last one, which is hard to integrate directly. We will find other way to get the analytical result.
In this section, at first we discuss some difficulties encountered in the calculation of the first 12 master integrals, and then concentrate on calculating the last master integral.

For the 4th to 10th master integrals, after integrating over $k_{2\perp}$, there is still a term proportional to
\begin{equation}\label{eq:kuvdiv}
	\int \frac {\ud^{D-2} k_{1\perp}}{(2 \pi)^{D-2}}
		\frac{1}{(k_{1\perp}^{2}+a) (k_{1\perp}^{2}+b)^{\epsilon}} \, ,
\end{equation}
where $a$ and $b$ are both nonnegative functions of $z$ and $z_1$. This integral on the one hand is cumbersome to expand $\epsilon$ after the integration, and on the other hand is ultraviolet divergent and thus cannot expand $\epsilon$ at the integrand level. We rewrite Eq.~\eqref{eq:kuvdiv} as
\begin{equation}
	\int \frac {\ud^{D-2} k_{1\perp}}{(2 \pi)^{D-2}}
		\frac{b-a}{(k_{1\perp}^{2}+a) (k_{1\perp}^{2}+b)^{1+\epsilon}}
	+ \int \frac {\ud^{D-2} k_{1\perp}}{(2 \pi)^{D-2}}
		\frac{1}{(k_{1\perp}^{2}+b)^{1+\epsilon}} \, ,
\end{equation}
where the second term can be integrated and then expand $\epsilon$ easily, while the first term is ultraviolet finite and thus can expand $\epsilon$ at the integrand level. For the first term, we need to expand to second order in $\epsilon$ and thus results in one-dimensional integrals
\begin{equation}
	\int_{0}^{\infty} \ud x
	\frac{b-a}{(x+a)(x+b)}
	=
	\ln b - \ln a \, ,
\end{equation}
and
\begin{equation}
	\int_{0}^{\infty} \ud x
	\frac{(b-a) \left[ \ln x + \ln (x+b) \right]}
		{(x+a)(x+b)}
	=
	\mathrm{Li}_2 \left( 1-\frac{a}{b} \right)
	- \ln a \, \ln b
	- \frac{1}{2} \ln^{2} a
	+ \frac{3}{2} \ln^2 b \, .
\end{equation}
Then we can integrate over $z_1$ easily.%, and get final result as linear function of PolyLog ($\mathrm{Li}_2 ,\mathrm{Li}_3 $).

For the 11th master integral, after integrating over $k_{2\perp}$, the master integral is proportional to
\begin{align}
	& \int_0^1 \ud z_1 \int \frac {\ud^{D-2} k_{1\perp}}{(2 \pi)^{D-2}}
		\frac{z_1^{1+\epsilon} (1-z_1)^{-\epsilon} (1+a z_1)^{-1+2\epsilon} }{(k_{1\perp}^{2}+a^2 z_1^2)^2 (k_{1\perp}^{2}+a^2 z_1^2+a z_1)^{\epsilon}} \, ,
\end{align}
with $a=(1-z)/z$, which is infrared divergent when integrating over $z_1$ near the region $z_1=0$. Thus one cannot expand $\epsilon$ at the integrand level. Yet, we can re-scale $k_{1\perp}$ by a factor of $z_1$, and we get
\begin{equation}\label{eq:irint}
\int_0^1 \ud z_1 \, z_1^{-1-2\epsilon} \int \frac {\ud^{D-2} k_{1\perp}}{(2 \pi)^{D-2}}
		\frac{(1-z_1)^{-\epsilon} (1+a z_1)^{-1+2\epsilon} }{(k_{1\perp}^{2}+a^2)^2 (z_1 k_{1\perp}^{2}+a^2 z_1+a)^{\epsilon}}\, ,
\end{equation}
which although is still infrared divergent, but we can expand the integrand other than $z_1^{-1-2\epsilon}$ as a power series of $\epsilon$.

Now let's concentrate on the last master integral
\begin{align}
\int \ud \Phi \frac{1}{E_1 E_2 E_4} ,
\end{align}
which is hard to calculate using the traditional integration method with Feynman parametrization.
Yet we can calculate it by constructing and soloving a differential equation \cite{Kotikov:1990kg,Remiddi:1997ny,Argeri:2007up,Henn:2014qga}. We define
\begin{equation}
	g(z) = \int \ud \Phi \frac{z^2}{E_1 E_2 E_4}
		 = \int
		\frac{\mathrm{d}^{D} k}{(2\pi)^{D}}
        \frac{\mathrm{d}^{D} l}{(2\pi)^{D}}
        \left ( \frac{i}{2 \pi} \right )^{3}
        \frac{P \cdot n}{S}
        \frac{1}{E_1 E_2 E_4 E_5 E_6 E_7}
        + \ldots \, ,
\end{equation}
where we omit 7 other similar terms.
Denominators of it do not contain $z$ except $E_7$. If we take the derivative of $g(z)$, it becomes
\begin{equation}
	\frac{\ud g(z)}{\ud z} =
	\int
	\frac{\mathrm{d}^{D} k}{(2\pi)^{D}}
    \frac{\mathrm{d}^{D} l}{(2\pi)^{D}}
    \left ( \frac{i}{2 \pi} \right )^{3}
    \frac{- (P \cdot n)^2}{z^2 S}
    \frac{1}{E_1 E_2 E_4 E_5 E_6 E_7^2 }
    + \ldots \, .
\end{equation}
Then we can reduce the integrals again by using IBP and arrive at a differential equation about $g(z) $
\begin{equation}\label{eq:dfe}
	\frac{\ud g(z)}{\ud z} = \frac{2 (z-1) z }{2 z - 1} \epsilon \, g(z) + h(z) \, ,
\end{equation}
where $h(z)$ is a linear combination of the first 12 master integrals, which gives
\begin{equation}
	h(z) = \frac{ (\ln z - \ln (1-z) )^2}{128 \pi ^4 (1-2 z)} \, .
\end{equation}
It is easy to see that $g(z)$ has no divergence, and thus the term proportional to $\epsilon$ in Eq.~\eqref{eq:dfe} can be safely omitted. Thus the differential equation can be solved by integrating $h(z)$ over $z$ combined with an initial value. A good choice of the initial value can be at $z=1$, where one gets $g(1) = 0$ as the integral over plus direction is suppressed.
With this initial value, we eventually get
\begin{align}\label{eq:mi13}
  \begin{autobreak}
%  \MoveEqLeft
    I_{13} = - \frac{1}{128 \pi ^4 z^2}
       \Bigg(  \mathrm{Li}_3 \left( \frac{2 z-1}{z} \right)
      + \mathrm{Li}_3 \left( \frac{z}{z-1} \right)
      + \mathrm{Li}_3 \left( \frac{2 z - 1}{z-1} \right)
      - \mathrm{Li}_2 (z) \ln \left( \frac{1-z}{z} \right)
      + \mathrm{Li}_2 \left( \frac{2 z-1}{z-1} \right) \ln \left( \frac{1-z}{z} \right)
      + \frac{\ln^3 \left( \frac{1-z}{z} \right)}{6}
      - \frac{\ln z \, \ln (1-z) \, \ln \left( \frac{1-z}{z} \right)}{2}
      - \zeta (3) \Bigg)
  \end{autobreak}.
\end{align}

\subsection{Analytical results}

Substituting analytical results for the thirteen master integrals into Eq.~\eqref{eq:sdcmi}, we find all kinds of divergences are canceled, and finite result gives
\begin{equation}\label{eq:d0}
	d^{(0)} (z,M,\mu_0) =
	\frac{128 (N_c^2 - 4) \pi^3 \alpha_s^3}{3 N_c^2 M^3}
	\left( C I_{13}
		+\sum_{i=0}^{11} C_i \, L_i
	\right) \, ,
\end{equation}
where $I_{13}$ is given in Eq.~\eqref{eq:mi13}, coefficients $C$ and $C_i(i=0,\ldots,11)$ are given in Eq.~\eqref{eq:d0coeff} in the Appendix, and $L_i(i=0,\ldots,11)$ are defined as
\begin{align}\label{eq:d0logs}
\begin{split}
&	L_0 = 1 \, , \
	L_1 = \ln z \, , \
	L_2 = \ln (1-z) \, , \
	L_3 = \ln (2-z) \, , \
	L_4 = \ln^2 z \, , \
	L_5 = \ln^2 (1-z) \, , \
	L_6 = \ln^2 (2-z) \, ,  \\
&	L_7 = \ln z \, \ln (1-z) \, , \
	L_8 = \ln z \, \ln (2-z) \, , \
	L_9 = \li_2 (1-z) \, , \
	L_{10} = \li_2 \left(\frac{z-1}{z-2}\right) \, , \
	L_{11} = \li_2 \left(\frac{2 (z-1)}{z-2}\right) \, .
\end{split}
\end{align}

For the transversely polarized SDC $d_T^{(0)} (z,M,\mu_0)$, we can express it similar to $d^{(0)} (z,M,\mu_0)$ in Eq.~\eqref{eq:d0}, but with different coefficients $C^T$ and $C^T_i(i=0,\ldots,11)$ given in Eq.~\eqref{eq:d0Tcoeff}. The relativistic correction SDC $d^{(2)}(z,M,\mu_0)$ and corresponding transverse polarized SDC $d_T^{(2)}(z,M,\mu_0)$ can also be expressed the same as  that in Eq.~\eqref{eq:d0} with corresponding coefficients given in Eq.~\eqref{eq:d2coeff} and Eq.~\eqref{eq:d2Tcoeff}.

As we note in Sec. \ref{sec:def}, LO SDCs $d^O(z,2m_Q,\mu_0)$ (similar for $d_T^O(z,2m_Q,\mu_0)$) in NRQCD factorization can be obtained by replacing $M$ in Eq.~\eqref{eq:d0} by $2m_Q$ and keeping other coefficients unchanged. Coefficients of the relativistic correction SDCs $d^P(z,2m_Q,\mu_0)$ and $d_T^P(z,2m_Q,\mu_0)$ are given in Eq.~\eqref{eq:dPcoeff} and Eq.~\eqref{eq:dPTcoeff}.

\subsection{\texorpdfstring{Large $z$ behaviour}{Large z behaviour}}

At hadron colliders, high $p_T$ quarkonium production is most sensitive to fragmentation function at large $z$ region. Thus we investigate SDCs obtained above at this region by expanding them around $z\to1$, and we get
\begin{align}\label{eq:largez}
  \begin{autobreak}
    d^{(0)} (z,M,\mu_0) = \frac{4 (N_c^2 - 4) \alpha_s^3}{3 \pi N_c^2 M^3}
                  \bigg((1-z) \big(-\ln (1-z)-3 \big)
                  +\frac{(1-z)^2}{18} \big(36 \ln ^2(1-z)
                  +18 \ln (1-z)
                  +4 \pi ^2
                  +93\big)
                  +O\big((1-z)^3\big) \bigg) \, ,
  \end{autobreak}\notag\\
  \begin{autobreak}
    d_T^{(0)} (z,M,\mu_0) = \frac{4 (N_c^2 - 4) \alpha_s^3}{3 \pi N_c^2 M^3}
                    \bigg((1-z) \big(-\ln (1-z)-3 \big)
                    +\frac{(1-z)^2}{9} \big(18 \ln ^2(1-z)
                    +18 \ln (1-z)
                    +2 \pi ^2
                    +51\big)
                    +O\big((1-z)^3\big) \bigg) \, ,
  \end{autobreak}\notag\\
  \begin{autobreak}
    d^{(2)} (z,M,\mu_0) = \frac{4 (N_c^2 - 4) \alpha_s^3}{3 \pi N_c^2 M^3}
                  \bigg(\frac{2}{135} \big(-15+22 \pi^2\big)
                  +\frac{1-z}{27} \big(-63 \ln ^2(1-z)
                  -81 \ln (1-z)
                  -239\big)
                  +O\big((1-z)^2\big) \bigg) \, ,
  \end{autobreak}\notag\\
  \begin{autobreak}
    d_T^{(2)} (z,M,\mu_0) = \frac{4 (N_c^2 - 4) \alpha_s^3}{3 \pi N_c^2 M^3}
                    \bigg(\frac{2}{135} \big(-15+22 \pi ^2\big)
                    +\frac{1-z}{27} \big(-63 \ln ^2(1-z)
                    -81 \ln (1-z)
                    -257\big)
                    +O\big((1-z)^2\big) \bigg) \, ,
  \end{autobreak}\notag\\
  \begin{autobreak}
    d^P (z,2m_Q,\mu_0) = \frac{4 (N_c^2 - 4) \alpha_s^3}{3 \pi N_c^2 (2m_Q)^3}
                  \bigg(\frac{2}{135} \big(-15+22 \pi^2\big)
                  +\frac{1-z}{54} \big(-126 \ln ^2(1-z)
                  -81 \ln (1-z)
                  -235\big)
                  +O\big((1-z)^2\big) \bigg) \, ,
  \end{autobreak}\notag\\
  \begin{autobreak}
    d_T^P (z,2m_Q,\mu_0) = \frac{4 (N_c^2 - 4) \alpha_s^3}{3 \pi N_c^2 (2m_Q)^3}
                    \bigg(\frac{2}{135} \big(-15+22 \pi ^2\big)
                    +\frac{1-z}{54} \big(-126 \ln ^2(1-z)
                    -81 \ln (1-z)
                    -271\big)
                    +O\big((1-z)^2\big) \bigg) \, .
  \end{autobreak}
\end{align}
We find that, for all cases, polarization-summed SDC equals to transversely polarized SDC at lowest order in $1-z$ expansion, while there are differences at higher orders. Thus longitudinal polarized SDCs are negligible at large $z$ region. The physical reason is very simple. As the two final state gluons are very soft when $z\to1$, heavy quark spin symmetry ensures that soft gluon emission will not change the spin of heavy quark. Therefore, the finial state heavy quark pair has almost the same polarization as that of the fragmenting gluon, which is transversely polarized.  The consequence is that, for high $p_T$ quarkonium production, contributions from gluon fragmentating to $\CScSa$ channel are transversely polarized, for both SGF and NRQCD factorization.

Another information from Eq.~\eqref{eq:largez} is that, at large $z$ region, relativistic correction terms are much larger than corresponding lowest order terms. This is because nonrelativistic expansion enhances the power of heavy quark propagator denominators, which vanish as $z\to1$. If fact, there are even infrared divergences if one expands to $O(v^4)$ terms  \cite{Bodwin:2012xc}, and the divergences need to be removed by color-octet mechanism. Based on this, it makes no sense to compare the convergence of velocity expansion between SGF and NRQCD factorization for the current problem.

\begin{figure}[htb!]
 \begin{minipage}[h]{0.75\linewidth}
 \begin{center}
	 \includegraphics[width=0.95\textwidth]{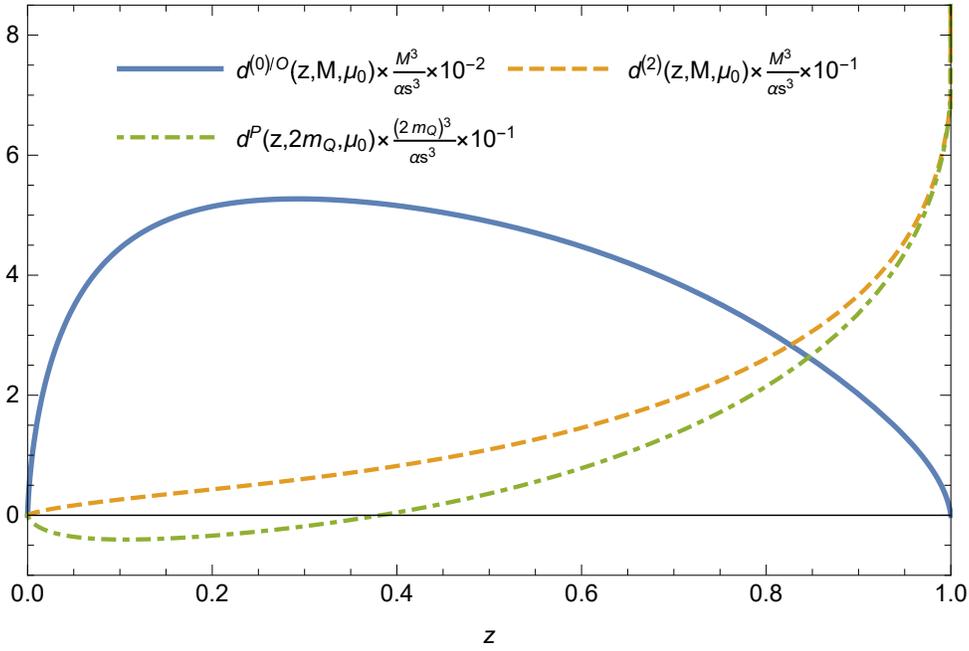}
	  \caption{Polarization-summed SDCs as functions of $z$. Solid curve corresponds to lowest order in $v^2$ expansion in either SGF (with superscript ``$(0)$") or NRQCD (with superscript ``$O$"), based on the relation in Eq.~\eqref{eq:relationU}. Dashed curve corresponds to order $v^2$ expansion in SGF.  Dash-dotted curve corresponds to order $v^2$ expansion in NRQCD. }
\label{fig:cpall}
 \end{center}
 \end{minipage}
 \end{figure}

 \begin{figure}[htb!]\label{fig:cpallT}
 \begin{minipage}[h]{0.75\linewidth}
 \begin{center}
	 \includegraphics[width=0.95\textwidth]{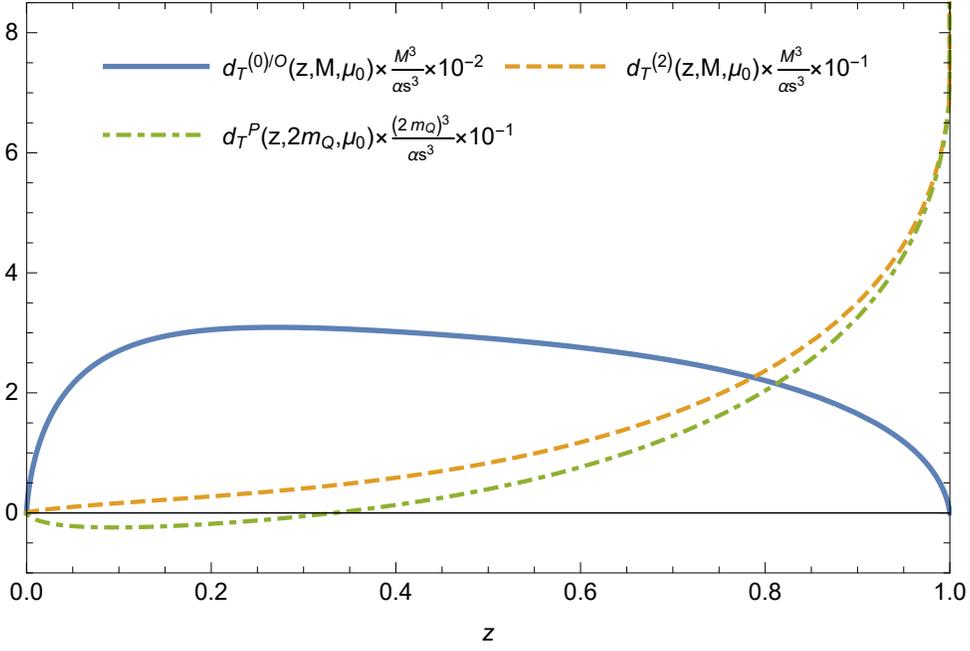}
	  \caption{Transversely polarized SDCs as functions of $z$. Meaning of each curve is similar to that in Fig.~\ref{fig:cpall}.}
\label{fig:cpallT}
 \end{center}
 \end{minipage}
\end{figure}
%
%Compairing the results between the SGF and NRQCD as shown in Fig.~\ref{fig:cpall} and Fig.~\ref{fig:cpallT}, we find that the LO FFs in SGF are exactly depressed by $\lambda^3$ where we denote $\lambda = (2m_Q)/M$. However, the relative-$v^2$ correction FFs in SGF are entirely depressed by a factor of $\lambda^5$, which implies that the the $v$ expansion in SGF in more proper than that in NRQCD as we expect in Ref.~\cite{Ma:2017xno}.

\begin{figure}[htb!]
 \begin{minipage}[h]{0.75\linewidth}
 \begin{center}
	 \includegraphics[width=0.95\textwidth]{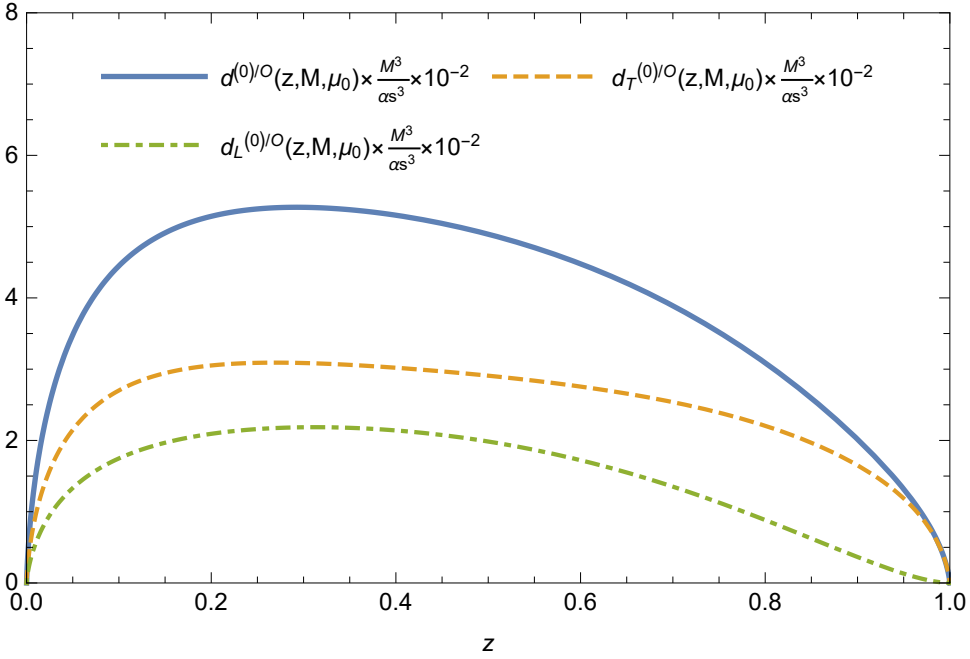}
	  \caption{SDCs at lowest order in $v^2$ expansion in either SGF (with superscript ``$(0)$") or NRQCD (with superscript ``$O$"), based on relations in Eqs.~\eqref{eq:relationU} and \eqref{eq:relationT}. Solid curve represents polarization summed SDCs, dashed curve represents transversely polarized SDCs, and dash-dotted curve represents longitudinally polarized SDCs.  \label{fig:cpLOTL}}
 \end{center}
 \end{minipage}
  \end{figure}

 \begin{figure}[htb!]
 \begin{minipage}[h]{0.75\linewidth}
 \begin{center}
	 \includegraphics[width=0.95\textwidth]{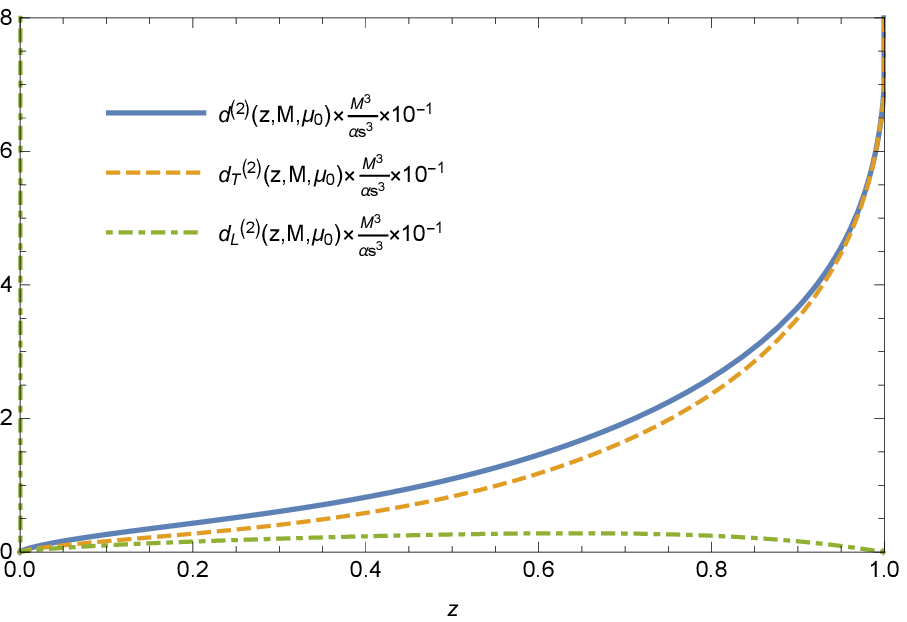}
	  \caption{SDCs at order $v^2$ expansion in SGF framework.  Meaning of each curve is similar to that in Fig.~\ref{fig:cpLOTL}. \label{fig:cpSGFTL}}
 \end{center}
 \end{minipage}
  \end{figure}

 \begin{figure}[htb!]
 \begin{minipage}[h]{0.75\linewidth}
 \begin{center}
	 \includegraphics[width=0.95\textwidth]{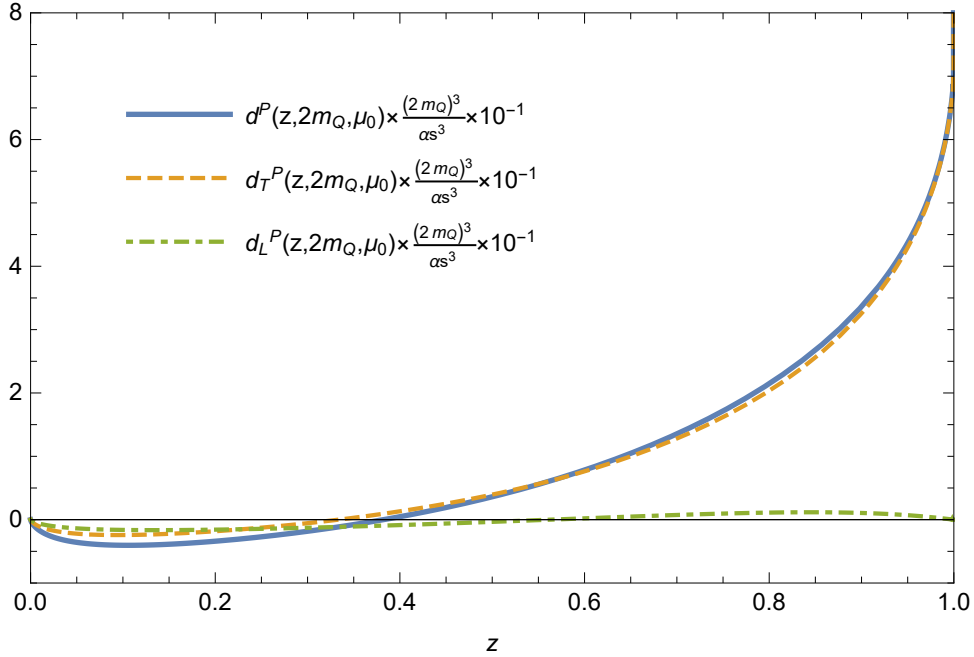}
	  \caption{SDCs at order $v^2$ expansion in NRQCD framework.  Meaning of each curve is similar to that in Fig.~\ref{fig:cpLOTL}. \label{fig:cpNRQCDTL}}
 \end{center}
 \end{minipage}
\end{figure}

\section{Numerical results and discussion} \label{sec:summary}

We plot our polarization-summed and transversely polarized SDCs in Fig.~\ref{fig:cpall} and Fig.~\ref{fig:cpallT}, respectively. We find that our $d^O(z,2m_Q,\mu_0)$ is compatible with the numerical result in Refs. \cite{Braaten:1993rw,Braaten:1995cj,Bodwin:2003wh}, and $d^P(z,2m_Q,\mu_0)$ is compatible with the numerical result in Ref. \cite{Bodwin:2003wh}. Our polarized SDC $d_T^O(z,2m_Q,\mu_0)$ seems to be not compatible with the result extracted from physical cross section in Ref.\cite{Qi:2007sf}. Other results calculated in this paper are new.

In Fig.~\ref{fig:cpLOTL}, Fig.~\ref{fig:cpSGFTL} and Fig.~\ref{fig:cpNRQCDTL}, we compare polarization-summed, transversely polarized, and longitudinal polarized SDCs for each case. As expected, polarization-summed SDCs approach transversely polarized SDCs as $z\to1$.

\begin{table}[htb!]
\begin{tabular}{|c|c|c|c|c|c|c|c|c|c|c|}
	\hline
	\multirow{2}{*}{Framework} & \multirow{2}{*}{Polarization} & \multicolumn{3}{c|}{$z^2$} & \multicolumn{3}{c|}{$z^4$} & \multicolumn{3}{c|}{$z^6$} \\
	\cline{3-11}
	 & & $F$ & $c_1$ & $c_2$ & $F$ & $c_1$ & $c_2$ & $F$ & $c_1$ & $c_2$ \\
	\hline
	\multirow{2}{*}{SGF} & Sum & \multirow{4}{*}{$1.28\times 10^{-3}$} & 1 & 9.07 & \multirow{4}{*}{$6.15\times 10^{-4}$} & 1 & 13.9 & \multirow{4}{*}{$3.71\times 10^{-4}$} & 1 & 18.4  \\
	\cline{2-2} \cline{4-5} \cline{7-8} \cline{10-11}
	 & Transvers & & 0.679 & 8.42 & & 0.724 & 13.2 & & 0.759 & 17.7 \\
	\cline{1-2} \cline{4-5} \cline{7-8} \cline{10-11}
	\multirow{2}{*}{NRQCD} & Sum & & 1 & 7.58 & & 1 & 12.4 & & 1 & 16.9 \\
	\cline{2-2} \cline{4-5} \cline{7-8} \cline{10-11}
	 & Transvers & & 0.679 & 7.41 & & 0.724 & 12.1 & & 0.759 & 16.5 \\
	\hline
\end{tabular}
\caption{Coefficients to estimate relative importance of each part of FFs calculated in either SGF or NRQCD framework. \label{table:FFpro}}
\end{table}

%We also find that the relativistic corrections in NRQCD may be negative at small $z$, but it does not matter as the fragmentation probability is obtained by integrating the FF over the longitudinal fraction $z$.
To estimate the relative contribution of each term for cross section, we integrate FFs calculated in this paper with a test function,
\begin{align}\label{eq:numff}
\begin{split}
  \int_0^1 \ud z \, z^{n} \, D^{\textrm{SGF}}_{g\to H} (z)
  & = F \cdot \lambda^3
      \frac{\alpha_s^3}{m_Q^3} \LDME{H}{\CScSa}
      (c_1+c_2 \lambda^2 v^2+O(v^4) ) \, ,  \\
  \int_0^1 \ud z \, z^{n} \, D^{\textrm{NRQCD}}_{g\to H} (z)
  & =  F \cdot
      \frac{\alpha_s^3}{m_Q^3} \LDME{H}{\CScSa}
      (c_1+c_2 v^2+O(v^4) ) \, ,
\end{split}
\end{align}
where we denote $\lambda = m_Q/E$ and $v^2=E^2/m_Q^2-1$. The factors $F$, $c_1$ and $c_2$ depend on $n$, polarization and factorization method. For $n=2, 4, 6$, corresponding factors are shown in the Table.~\ref{table:FFpro}. With larger $n$, the integration in Eq.~\eqref{eq:numff} probes larger $z$, we then find $c_2/c_1$ also becomes larger which is consistent with our observation of large $z$ behaviour.

%Then we can estimate the relative sizes for the fragmentation if we provide the value of $v^2$. We can take the $\overline{MS}$ mass of the charm quark as $m_c=1.27 \gev$, corresponding to the latest Particle Data Group \cite{Olive:2016xmw}, and the heavy quark pair mass $M$ can be taken $3.1\gev$ or $3.5\gev$ to make some tests, as the charmonium mass $M_{\jpsi}=3.097 \gev$. Then if we take $M=3.1\gev$, we can get $\lambda=0.8$ and the corresponding $v^2=0.5$. If we take the $z^4$ integral for example, then the relativistic correction make the contributions by about 4 times and 6 times with respect to the LO in SGF and NRQCD separately. And if we take $M=3.1\gev$, then $\lambda=0.7$ and $v^2=0.9$. The relativistic contributions becomes 6 times and 11 times respectively. We can find that the SGF make sense in the depression of the relativitic corrections.
%
%
%In addition, the transverse contribution is almost twice of the longitudinal contribution for the LO of $v$, while the transverse quarkonium is dominant in the relativistic corrections as the longitudinal contribution is almost zero through $z$.

%--------------------------------------------------------------------
\begin{acknowledgments}
%--------------------------------------------------------------------

We thank Haoyu Liu, Ce Meng, Chenyu Wang, Yujie Zhang and Huaxing Zhu for many useful communications and discussions.
The work is supported in part by the National Natural Science Foundation
of China (Grants No. 11475005 and No. 11075002), and the National Key Basic Research Program
of China (No. 2015CB856700).

%--------------------------------------------------------------------
\end{acknowledgments}
%--------------------------------------------------------------------

\appendix*

\section{\textbf{Coefficients}}\label{ap:coeff}
%All the SDCs in the FFs of gluon to $\CScSa$ have the form in Eq.~\eqref{eq:d0}, where $I_{13}$ is given in Eq.~\eqref{eq:mi13} and we will give the coefficients $C$ and $C_i(i=0,\ldots,11)$ here.

Coefficients $C$ and $C_i(i=0,\ldots,11)$  of SDCs $d^{(0)}(z,M,\mu_0)$ and $d^O(z,2m_Q,\mu_0)$ defined in Eq.~\eqref{eq:d0} are
\begin{align}\label{eq:d0coeff}
  C   & = \frac{(4 z+1) z^2}{4} \, , \notag\\
  C_0 & = \frac{32 \pi ^2 z^4+330 z^4-112 \pi ^2 z^3-861 z^3+144 \pi ^2 z^2+876 z^2-80 \pi ^2 z-273 z+16 \pi ^2}{2304 \pi ^4 (z-1) (2 z-1)} \, , \notag \\
  C_1 & = -\frac{z \left(8 z^6+36 z^5-338 z^4+741 z^3-599 z^2+195 z-19\right)}{768 \pi ^4 (z-1)^2 (2 z-1)^2} \, , \notag \\
  C_2 & = -\frac{(z-1) \left(8 z^4-76 z^3+10 z^2+73 z-24\right)}{768 \pi ^4 (2 z-1)^2} \, , \notag \\
  C_3 & = \frac{(z-2) z \left(z^3-z^2-7 z+1\right)}{192 \pi ^4 (z-1)^2} \, , \notag \\
  C_4 & = -\frac{z \left(32 z^7-160 z^6+434 z^5-763 z^4+734 z^3-376 z^2+96 z-9\right)}{1536 \pi ^4 (z-1)^3 (2 z-1)^3}  \, , \notag \\
  C_5 & = -\frac{(z-1) \left(16 z^6-168 z^5-12 z^4+474 z^3-528 z^2+243 z-40\right)}{1536 \pi ^4 (2 z-1)^3} \, , \notag \\
  C_6 & = \frac{z^7-13 z^6+25 z^5+43 z^4-206 z^3+282 z^2-166 z+40}{768 \pi ^4 (z-1)^3} \, , \notag \\
  C_7 & = \frac{(z-1) z \left(16 z^5-168 z^4-12 z^3+154 z^2-48 z+3\right)}{768 \pi ^4 (2 z-1)^3} \, , \notag \\
  C_8 & = -\frac{z \left(z^6-13 z^5+25 z^4+3 z^3-46 z^2+42 z-6\right)}{384 \pi ^4 (z-1)^3} \, , \notag \\
  C_9 & = \frac{5 (z-1)}{96 \pi ^4} \, , \notag \\
  C_{10} & = \frac{5 (z-1)}{48 \pi ^4} \, , \notag \\
  C_{11} & = -\frac{(z-1) (2 z+3)}{48 \pi ^4}  \, . 
\end{align}

Coefficients of transverse polarized SDCs $d_T^{(0)}(z,M,\mu_0)$ and $d_T^O(z,2m_Q,\mu_0)$ are
\begin{align}\label{eq:d0Tcoeff}
  \begin{autobreak}
    \MoveEqLeft
    C^T   = \frac{1}{32}
            (40 z^3
            -32 z^2
            +96 z
            -79) \, ,
  \end{autobreak}\notag\\
  \begin{autobreak}
    \MoveEqLeft
    C^T_0 = \frac{1}{645120 \pi ^4 (z-1) z^2 (2 z-1)}
            (1296 z^8
            +9288 z^7
            +8960 \pi ^2 z^6
            +57948 z^6
            -40320 \pi ^2 z^5
            -213318 z^5
            +80640 \pi ^2 z^4
            +345048 z^4
            -94080 \pi ^2 z^3
            -321345 z^3
            +67200 \pi ^2 z^2
            +186918 z^2
            -26880 \pi ^2 z
            -45675 z
            +4480 \pi ^2) \, ,
  \end{autobreak}\notag\\
  \begin{autobreak}
    \MoveEqLeft
    C^T_1 = -\frac{1}{215040 \pi ^4 (z-1)^2 z (2 z-1)^2}
            (576 z^{10}
            +2784 z^9
            -17392 z^8
            +44344 z^7
            -129280 z^6
            +284936 z^5
            -300650 z^4
            +141647 z^3
            -9049 z^2
            -15011 z
            +3815) \, ,
  \end{autobreak}\notag\\
  \begin{autobreak}
    \MoveEqLeft
    C^T_2 = -\frac{z-1}{215040 \pi ^4 z^2 (2 z-1)^2}
            (576 z^8
            +3936 z^7
            -10384 z^6
            -9160 z^5
            -17288 z^4
            +66856 z^3
            -112526 z^2
            +81671 z
            -19040) \, ,
  \end{autobreak}\notag\\
  \begin{autobreak}
    \MoveEqLeft
    C^T_3 = \frac{1}{6720 \pi ^4 (z-1)^2 z}
            (9 z^8
            +48 z^7
            -250 z^6
            +344 z^5
            -693 z^4
            +1716 z^3
            -1708 z^2
            +832 z
            -88) \, ,
  \end{autobreak}\notag\\
  \begin{autobreak}
    \MoveEqLeft
    C^T_4 = -\frac{1}{12288 \pi ^4 (z-1)^3 z (2 z-1)^3}
            (304 z^9
            -1560 z^8
            +5212 z^7
            -13046 z^6
            +21700 z^5
           -24425 z^4
           +18240 z^3
           -8468 z^2
           +2184 z
           -237) \, ,
  \end{autobreak}\notag\\
  \begin{autobreak}
    \MoveEqLeft
    C^T_5 = -\frac{z-1}{430080 \pi ^4 z^2 (2 z-1)^3}
            (1152 z^{10}
            +6720 z^9
            -28448 z^8
            +3760 z^7
            -16080 z^6
            +84440 z^5
            -105484 z^4
            +64094 z^3
            -8324 z^2
            -6119 z
            +1664) \, ,
  \end{autobreak}\notag\\
  \begin{autobreak}
    \MoveEqLeft
    C^T_6 = \frac{1}{26880 \pi ^4 (z-1)^3 z^2}
            (9 z^{11}
            +30 z^{10}
            -340 z^9
            +666 z^8
            -603 z^7
            +1876 z^6
            -6076 z^5
            +10020 z^4
            -9132 z^3
            +4656 z^2
            -688 z
            -208) \, ,
  \end{autobreak}\notag\\
  \begin{autobreak}
    \MoveEqLeft
    C^T_7 = \frac{1}{215040 \pi ^4 z (2 z-1)^3}
            (1152 z^{10}
            +5568 z^9
            -35168 z^8
            +32208 z^7
            -19840 z^6
            +21672 z^5
            +14364 z^4
            -102550 z^3
            +114590 z^2
            -51555 z
            +8295) \, ,
  \end{autobreak}\notag\\
  \begin{autobreak}
    \MoveEqLeft
    C^T_8 = -\frac{z}{13440 \pi ^4 (z-1)^3}
            (9 z^8
            +30 z^7
            -340 z^6
            +666 z^5
            -603 z^4
            +644 z^3
            -1036 z^2
            +980 z
            -140) \, ,
  \end{autobreak}\notag\\
  \begin{autobreak}
   \MoveEqLeft
    C^T_9 = \frac{1}{1680 \pi ^4 z^2}
            (77 z^3
            -84 z^2
            +82 z
            +13) \, ,
  \end{autobreak}\notag\\
  \begin{autobreak}
    \MoveEqLeft
    C^T_{10} = \frac{1}{840 \pi ^4 z^2}
              (77 z^3
              -84 z^2
              +82 z
              +13) \, ,
  \end{autobreak}\notag\\
  \begin{autobreak}
    \MoveEqLeft
    C^T_{11} = -\frac{1}{840 \pi ^4 z^2}
              (35 z^4
              -28 z^3
              +56 z^2
              -23 z
              +48)  \, .
  \end{autobreak}
\end{align}

Coefficients of SDC $d^{(2)}(z,M,\mu_0)$ in SGF are
\begin{align}\label{eq:d2coeff}
  \begin{autobreak}
    \MoveEqLeft
    C^{(2)}   =  \frac{z^2 (3 z-5)}{2}  \, ,
  \end{autobreak}\notag\\
  \begin{autobreak}
    \MoveEqLeft
    C^{(2)}_0   =  -\frac{1}{17280 \pi ^4 (z-2)^2 (z-1)^3 (2 z-1)^3}
    				(-6264 z^{11}
    				+32 \pi ^2 z^{10}
    				+66528 z^{10}
    				-336 \pi ^2 z^9
    				-308028 z^9
    				+136 \pi ^2 z^8
    				+820695 z^8
    				+7876 \pi ^2 z^7
    				-1393095 z^7
    				-35660 \pi ^2 z^6
    				+1574844z^6
    				+75048 \pi ^2 z^5
    				-1207038 z^5
    				-90608 \pi ^2 z^4
    				+625245 z^4
    				+66084 \pi ^2 z^3
    				-211563 z^3
    				-28732 \pi ^2 z^2
    				+42708 z^2
    				+6848 \pi ^2 z
    				-3996 z
    				-688 \pi^2) \, ,
  \end{autobreak} \notag\\
  \begin{autobreak}
    \MoveEqLeft
    C^{(2)}_1   =  -\frac{z}{5760 \pi ^4 (z-2)^2 (z-1)^4 (2 z-1)^4}
    				(2784 z^{13}
    				-32224 z^{12}
    				+161096 z^{11}
    				-454480 z^{10}
    				+789646 z^9
    				-851497 z^8
    				+507690 z^7
    				-39713 z^6
    				-199724 z^5
    				+176257 z^4
    				-75580 z^3
    				+17581z^2
    				-1860 z
    				+12) \, ,
  \end{autobreak} \notag\\
  \begin{autobreak}
    \MoveEqLeft
    C^{(2)}_2   =  -\frac{z-1}{5760 \pi ^4 (2 z-1)^4}
    				(2784 z^7
    				-9952 z^6
    				+6488 z^5
    				+8184 z^4
    				-14698 z^3
    				+8669 z^2
    				-2451z
    				+270) \, ,
  \end{autobreak} \notag\\
  \begin{autobreak}
    \MoveEqLeft
    C^{(2)}_3   =  \frac{z}{2880 \pi^4 (z-2)^2 (z-1)^4}
    				(174 z^9
    				-1753 z^8
    				+7227 z^7
    				-15310 z^6
    				+16211 z^5
    				-3641 z^4
    				-10826 z^3
    				+12860 z^2
    				-5896 z
    				+948) \, ,
  \end{autobreak} \notag\\
  \begin{autobreak}
    \MoveEqLeft
    C^{(2)}_4   =  -\frac{z}{11520 \pi ^4 (z-1)^5 (2 z-1)^5}
    				(2832 z^{11}
    				-26064 z^{10}
    				+109694 z^9
    				-278685 z^8
    				+468260 z^7
    				-541351 z^6
    				+438362 z^5
    				-248782 z^4
    				+97035 z^3
    				-24775 z^2
    				+3735z
    				-255) \, ,
  \end{autobreak} \notag\\
  \begin{autobreak}
    \MoveEqLeft
    C^{(2)}_5   =  -\frac{z-1}{11520 \pi ^4 (2 z-1)^5}
    				(5568 z^9
    				-25472 z^8
    				+42768 z^7
    				-33152 z^6
    				-2972 z^5
    				+32926 z^4
    				-32114 z^3
    				+14871 z^2
    				-3509z
    				+336) \, ,
  \end{autobreak} \notag\\
  \begin{autobreak}
    \MoveEqLeft
    C^{(2)}_6   =  \frac{1}{11520 \pi ^4(z-1)^5}
    				(174 z^{10}
    				-1405 z^9
    				+4775 z^8
    				-8896 z^7
    				+9267 z^6
    				-2989 z^5
    				-6460 z^4
    				+11700 z^3
    				-9350 z^2
    				+3850 z
    				-672) \, ,
  \end{autobreak} \notag\\
  \begin{autobreak}
    \MoveEqLeft
    C^{(2)}_7   =  \frac{z}{5760 \pi ^4 (2 z-1)^5}
    				(5568 z^9
    				-31040 z^8
    				+68240 z^7
    				-75920 z^6
    				+38500 z^5
    				+4346 z^4
    				-17360 z^3
    				+9705 z^2
    				-2340 z
    				+225) \, ,
  \end{autobreak} \notag\\
  \begin{autobreak}
    \MoveEqLeft
    C^{(2)}_8   =  -\frac{z}{5760 \pi ^4 (z-1)^5}
    				(174 z^9
    				-1405 z^8
    				+4775 z^7
    				-8896 z^6
    				+9787 z^5
    				-6261 z^4
    				+2100 z^3
    				-220 z^2
    				-30 z
    				-30) \, ,
  \end{autobreak} \notag\\
  \begin{autobreak}
    \MoveEqLeft
    C^{(2)}_9   =  -\frac{65 z-84}{1440 \pi ^4} \, ,
  \end{autobreak} \notag\\
  \begin{autobreak}
    \MoveEqLeft
    C^{(2)}_{10}   =  -\frac{65z-84}{720 \pi ^4} \, ,
  \end{autobreak} \notag\\
  \begin{autobreak}
    \MoveEqLeft
    C^{(2)}_{11}   =  \frac{z^2+63 z-127}{720 \pi ^4} \, .
  \end{autobreak} 
\end{align}

Coefficients of transverse polarized SDC $d_T^{(2)}(z,M,\mu_0)$ in SGF are
\begin{align}\label{eq:d2Tcoeff}
  \begin{autobreak}
    \MoveEqLeft
    C^{(2)\,T}   =  \frac{64 z^3-168 z^2+136 z-71}{48} \, ,
  \end{autobreak}\notag\\
  \begin{autobreak}
    \MoveEqLeft
    C^{(2)\,T}_0   =  -\frac{1}{967680 \pi ^4 (z-2)^2 (z-1)^3 z^2 (2 z-1)^3}
    					(-69120 z^{15}
    					+460800 z^{14}
    					-1207584 z^{13}
    					+98560 \pi ^2 z^{12}
    					+2847312 z^{12}
    					-1339520 \pi ^2 z^{11}
    					-13468080 z^{11}
    					+8045632 \pi ^2 z^{10}
    					+55263268z^{10}
    					-28419552 \pi ^2 z^9
    					-146809262 z^9
    					+66135328 \pi ^2 z^8
    					+264859148 z^8
    					-107291744 \pi ^2 z^7
    					-339774121 z^7
    					+124709536 \pi ^2 z^6
    					+314920956z^6
    					-104648992 \pi ^2 z^5
    					-209118576 z^5
    					+62816096 \pi ^2 z^4
    					+96647240 z^4
    					-26238240 \pi ^2 z^3
    					-29399341 z^3
    					+7221536 \pi ^2 z^2
    					+5270356 z^2
    					-1173760 \pi ^2z
    					-420980 z
    					+85120 \pi ^2) \, ,
  \end{autobreak} \notag\\
  \begin{autobreak}
    \MoveEqLeft
    C^{(2)\,T}_1   =  -\frac{1}{967680 \pi ^4 (z-2) (z-1)^4 z (2 z-1)^4}
    					(92160 z^{16}
    					-645120 z^{15}
    					+1886592 z^{14}
    					-3520320 z^{13}
    					+5526528 z^{12}
    					-1461104 z^{11}
    					-38168600 z^{10}
    					+154594400 z^9
    					-338137300 z^8
    					+490545354 z^7
    					-502217199 z^6
    					+370179179 z^5
    					-195786076 z^4
    					+72504256 z^3
    					-17831711 z^2
    					+2613387 z
    					-172410) \, ,
  \end{autobreak} \notag\\
  \begin{autobreak}
    \MoveEqLeft
    C^{(2)\,T}_2   =  -\frac{z-1}{967680 \pi ^4 z^2 (2 z-1)^4}
    					(92160 z^{11}
    					-92160 z^{10}
    					-2688 z^9
    					-533568 z^8
    					+359616 z^7
    					+1262000 z^6
    					-2875304 z^5
    					+3019912 z^4
    					-2062052 z^3
    					+867514 z^2
    					-199619 z
    					+19040) \, ,
  \end{autobreak} \notag\\
  \begin{autobreak}
    \MoveEqLeft
    C^{(2)\,T}_3   =  \frac{1}{20160 \pi ^4 (z-2) (z-1)^4 z}
    					(240 z^{12}
    					-1320 z^{11}
    					+2693 z^{10}
    					-3768 z^9
    					+6807 z^8
    					-1162 z^7
    					-45157 z^6
    					+133224 z^5
    					-197565 z^4
    					+178220 z^3
    					-100834 z^2
    					+33760 z
    					-5096) \, ,
  \end{autobreak} \notag\\
  \begin{autobreak}
    \MoveEqLeft
    C^{(2)\,T}_4   =  -\frac{1}{92160 \pi ^4 (z-1)^5 z (2 z-1)^5}
    					(17120 z^{13}
    					-164960 z^{12}
    					+760616 z^{11}
    					-2204320 z^{10}
    					+4400860 z^9
    					-6361810 z^8
    					+6875560 z^7
    					-5640463 z^6
    					+3507010 z^5
    					-1625045 z^4
    					+542780 z^3
    					-123185 z^2
    					+16950 z
    					-1065) \, ,
  \end{autobreak} \notag\\
  \begin{autobreak}
    \MoveEqLeft
    C^{(2)\,T}_5   =  -\frac{z-1}{645120 \pi ^4 z^2 (2 z-1)^5}
    					(61440 z^{13}
    					-122880 z^{12}
    					+69888 z^{11}
    					-364160 z^{10}
    					+888704 z^9
    					-819616 z^8
    					-416016 z^7
    					+2338672 z^6
    					-3769616 z^5
    					+3725268 z^4
    					-2396422 z^3
    					+951968 z^2
    					-209663 z
    					+19488) \, ,
  \end{autobreak} \notag\\
  \begin{autobreak}
    \MoveEqLeft
    C^{(2)\,T}_6   =  \frac{1}{80640 \pi ^4 (z-1)^5 z^2}
    					(240 z^{14}
    					-1320 z^{13}
    					+2853 z^{12}
    					-4328 z^{11}
    					+9624 z^{10}
    					-21070 z^9
    					+25355 z^8
    					+1284 z^7
    					-67276 z^6
    					+146628 z^5
    					-186980 z^4
    					+159140 z^3
    					-90384 z^2
    					+31064 z
    					-4872) \, ,
  \end{autobreak} \notag\\
  \begin{autobreak}
    \MoveEqLeft
    C^{(2)\,T}_7   =  \frac{1}{322560 \pi ^4 z (2 z-1)^5}
    					(61440 z^{13}
    					-184320 z^{12}
    					+192768 z^{11}
    					-434048 z^{10}
    					+1252864 z^9
    					-1708320 z^8
    					+762000 z^7
    					+816256 z^6
    					-1748096 z^5
    					+1671908 z^4
    					-1002330 z^3
    					+379750 z^2
    					-81375 z
    					+7455) \, ,
  \end{autobreak} \notag\\
  \begin{autobreak}
    \MoveEqLeft
    C^{(2)\,T}_8   =  -\frac{z}{40320 \pi ^4 (z-1)^5}
    					(240 z^{11}
    					-1320 z^{10}
    					+2853 z^9
    					-4328 z^8
    					+9624 z^7
    					-21070 z^6
    					+28155 z^5
    					-20860 z^4
    					+8148 z^3
    					-1204 z^2
    					-140 z
    					-140) \, ,
  \end{autobreak} \notag\\
  \begin{autobreak}
    \MoveEqLeft
    C^{(2)\,T}_9   =  -\frac{350 z^3-1018 z^2+838 z-609}{10080 \pi ^4 z^2} \, ,
  \end{autobreak} \notag\\
  \begin{autobreak}
    \MoveEqLeft
    C^{(2)\,T}_{10}   =  -\frac{350 z^3-1018 z^2+838 z-609}{5040 \pi ^4 z^2} \, ,
  \end{autobreak} \notag\\
  \begin{autobreak}
    \MoveEqLeft
    C^{(2)\,T}_{11}   =  \frac{385 z^4-1610 z^3+2104 z^2-1682 z+56}{5040 \pi ^4 z^2} \, .
  \end{autobreak} 
\end{align}

Coefficients of SDC $d^P(z,2m_Q,\mu_0)$ in NRQCD are
\begin{align}\label{eq:dPcoeff}
  \begin{autobreak}
    \MoveEqLeft
    C^P   =  - \frac{23 z^2}{8}
  \end{autobreak}\notag\\
  \begin{autobreak}
    \MoveEqLeft
    C^P_0 = -\frac{1}{69120 \pi ^4 (z-2)^2 (z-1)^3 (2 z-1)^3}
                (-25056 z^{11}
                +5888 \pi ^2 z^{10}
                +325512 z^{10}
                -61824 \pi ^2 z^9
                -1802892 z^9
                +278464 \pi ^2 z^8
                +5668770 z^8
                -705056 \pi ^2 z^7
                -11282745 z^7
                +1105120 \pi ^2 z^6
                +14929116 z^6
                -1112448 \pi ^2 z^5
                -13351782 z^5
                +720448 \pi ^2 z^4
                +7982790 z^4
                -290784 \pi ^2 z^3
                -3050937 z^3
                +67232 \pi ^2 z^2
                +672492 z^2
                -7168 \pi ^2 z
                -65124 z
                +128 \pi ^2) \, ,
  \end{autobreak}\notag\\
  \begin{autobreak}
    \MoveEqLeft
    C^P_1 = -\frac{z}{23040 \pi ^4 (z-2)^2 (z-1)^4 (2 z-1)^4}
                (11136 z^{13}
                -130336 z^{12}
                +647984 z^{11}
                -1739440 z^{10}
                +2512744 z^9
                -1049338 z^8
                -3053565 z^7
                +6915463 z^6
                -7355666 z^5
                +4771048 z^4
                -1956745 z^3
                +489679 z^2
                -66480 z
                +3468) \, ,
  \end{autobreak}\notag\\
  \begin{autobreak}
    \MoveEqLeft
    C^P_2 = -\frac{z-1}{23040 \pi ^4 (2 z-1)^4}
                (11136 z^7
                -41248 z^6
                +41072 z^5
                +16896 z^4
                -66712 z^3
                +51686 z^2
                -17409 z
                +2160) \, ,
  \end{autobreak}\notag\\
  \begin{autobreak}
    \MoveEqLeft
    C^P_3 = \frac{z}{5760 \pi ^4 (z-2)^2 (z-1)^4}
                (348 z^9
                -3551 z^8
                +14859 z^7
                -31790 z^6
                +32692 z^5
                -2017 z^4
                -33127 z^3
                +35890 z^2
                -15572 z
                +2256) \, ,
  \end{autobreak}\notag\\
  \begin{autobreak}
    \MoveEqLeft
    C^P_4 = -\frac{z}{46080 \pi ^4 (z-1)^5 (2 z-1)^5}
                (5568 z^{11}
                -58176 z^{10}
                +255536 z^9
                -640800 z^8
                +1030370 z^7
                -1130629 z^6
                +878198 z^5
                -489193 z^4
                +192570 z^3
                -50995 z^2
                +8190 z
                -615) \, ,
  \end{autobreak}\notag\\
  \begin{autobreak}
    \MoveEqLeft
    C^P_5 = -\frac{z-1}{46080 \pi ^4 (2 z-1)^5}
                (22272 z^9
                -104768 z^8
                +204192 z^7
                -161408 z^6
                -91808 z^5
                +312604 z^4
                -288566 z^3
                +134184 z^2
                -32171 z
                +3144) \, ,
  \end{autobreak}\notag\\
  \begin{autobreak}
    \MoveEqLeft
    C^P_6 = \frac{1}{23040 \pi ^4 (z-1)^5}
                (348 z^{10}
                -2855 z^9
                +10225 z^8
                -20132 z^7
                +19434 z^6
                +6037 z^5
                -46085 z^4
                +65520 z^3
                -48130 z^2
                +18770 z
                -3144) \, ,
  \end{autobreak}\notag\\
  \begin{autobreak}
    \MoveEqLeft
    C^P_7 = \frac{z}{23040 \pi ^4 (2 z-1)^5}
                (22272 z^9
                -127040 z^8
                +308960 z^7
                -365600 z^6
                +160480 z^5
                +76604 z^4
                -122450 z^3
                +57630 z^2
                -12195 z
                +1035) \,
  \end{autobreak}\notag\\
  \begin{autobreak}
    \MoveEqLeft
    C^P_8 = -\frac{z}{11520 \pi ^4 (z-1)^5}
                (348 z^9
                -2855 z^8
                +10225 z^7
                -20132 z^6
                +22274 z^5
                -11307 z^4
                -1965 z^3
                +5680 z^2
                -2490 z
                +210) \, ,
  \end{autobreak}\notag\\
  \begin{autobreak}
    \MoveEqLeft
    C^P_9 = -\frac{1}{2880 \pi ^4}
                (355 z
                -393) \, ,
  \end{autobreak}\notag\\
  \begin{autobreak}
    \MoveEqLeft
    C^P_{10} = -\frac{1}{1440 \pi ^4}
                  (355 z
                  -393) \, ,
  \end{autobreak}\notag\\
  \begin{autobreak}
    \MoveEqLeft
    C^P_{11} = \frac{1}{1440 \pi ^4}
                  (92 z^2
                  +171 z
                  -389) \, .
  \end{autobreak}
\end{align}

Coefficients of transverse polarized SDC $d_T^P(z,2m_Q,\mu_0)$ in NRQCD are
\begin{align}\label{eq:dPTcoeff}
  \begin{autobreak}
    \MoveEqLeft
    C^{P\,T}   = \frac{1}{192}
                    (-104 z^3
                    -384 z^2
                    -320 z
                    +427) \, ,
  \end{autobreak}\notag\\
  \begin{autobreak}
    \MoveEqLeft
    C^{P\,T}_0 = -\frac{1}{3870720 \pi ^4 (z-2)^2 (z-1)^3 z^2 (2 z-1)^3}
                    (-276480 z^{15}
                    +1889856 z^{14}
                    -4822560 z^{13}
                    +716800 \pi ^2 z^{12}
                    +12032928 z^{12}
                    -9067520 \pi ^2 z^{11}
                    -70954464 z^{11}
                    +51455488 \pi ^2 z^{10}
                    +319426276 z^{10}
                    -173875968 \pi ^2 z^9
                    -882759338 z^9
                    +391226752 \pi ^2 z^8
                    +1627632260 z^8
                    -619396736 \pi ^2 z^7
                    -2121809899 z^7
                    +708461824 \pi ^2 z^6
                    +2002251756 z^6
                    -589431808 \pi ^2 z^5
                    -1363770480 z^5
                    +353193344 \pi ^2 z^4
                    +653097752 z^4
                    -148176000 \pi ^2 z^3
                    -207921895 z^3
                    +41183744 \pi ^2 z^2
                    +39320572 z^2
                    -6791680 \pi ^2 z
                    -3328220 z
                    +501760 \pi ^2) \, ,
  \end{autobreak}\notag\\
  \begin{autobreak}
    \MoveEqLeft
    C^{P\,T}_1 = -\frac{1}{3870720 \pi ^4 (z-2) (z-1)^4 z (2 z-1)^4}
                    (368640 z^{16}
                    -2642688 z^{15}
                    +7556736 z^{14}
                    -11275008 z^{13}
                    +5641728 z^{12}
                    +51641392 z^{11}
                    -313531112 z^{10}
                    +978432104 z^9
                    -1943050972 z^8
                    +2633953998 z^7
                    -2520394677 z^6
                    +1724536409 z^5
                    -839859748 z^4
                    +284061472 z^3
                    -63250469 z^2
                    +8303889 z
                    -483630) \, ,
  \end{autobreak}\notag\\
  \begin{autobreak}
    \MoveEqLeft
    C^{P\,T}_2 = -\frac{z-1}{3870720 \pi ^4 z^2 (2 z-1)^4}
                    (368640 z^{11}
                    -430848 z^{10}
                    -373632 z^9
                    -603264 z^8
                    +1200000 z^7
                    +6206192 z^6
                    -20341448 z^5
                    +31919680 z^4
                    -31026596 z^3
                    +17385046 z^2
                    -5059913 z
                    +590240) \, ,
  \end{autobreak}\notag\\
  \begin{autobreak}
    \MoveEqLeft
    C^{P\,T}_3 = \frac{1}{40320 \pi ^4 (z-2) (z-1)^4 z}
                    (480 z^{12}
                    -2721 z^{11}
                    +5278 z^{10}
                    -3963 z^9
                    -480 z^8
                    +28411 z^7
                    -150686 z^6
                    +380973 z^5
                    -553800 z^4
                    +494932 z^3
                    -273020 z^2
                    +86456 z
                    -11776) \, ,
  \end{autobreak}\notag\\
  \begin{autobreak}
    \MoveEqLeft
    C^{P\,T}_4 = -\frac{1}{368640 \pi ^4 (z-1)^5 z (2 z-1)^5}
                    (13760 z^{13}
                    -214880 z^{12}
                    +1084064 z^{11}
                    -2659840 z^{10}
                    +3168700 z^9
                    -223390 z^8
                    -5421920 z^7
                    +9546683 z^6
                    -9179450 z^5
                    +5699545 z^4
                    -2341660 z^3
                    +616645 z^2
                    -94470 z
                    +6405) \, ,
  \end{autobreak}\notag\\
  \begin{autobreak}
    \MoveEqLeft
    C^{P\,T}_5 = -\frac{z-1}{2580480 \pi ^4 z^2 (2 z-1)^5}
                    (245760 z^{13}
                    -532992 z^{12}
                    +79104 z^{11}
                    -200960 z^{10}
                    +2334848 z^9
                    -2308192 z^8
                    -5316624 z^7
                    +16336672 z^6
                    -21943232 z^5
                    +18457476 z^4
                    -10241914 z^3
                    +3602600 z^2
                    -723677 z
                    +62976) \, ,
  \end{autobreak}\notag\\
  \begin{autobreak}
    \MoveEqLeft
    C^{P\,T}_6 = \frac{1}{161280 \pi ^4 (z-1)^5 z^2}
                    (480 z^{14}
                    -2721 z^{13}
                    +5598 z^{12}
                    -5137 z^{11}
                    +6864 z^{10}
                    -21665 z^9
                    +16978 z^8
                    +96447 z^7
                    -350984 z^6
                    +610488 z^5
                    -670420 z^4
                    +490468 z^3
                    -233184 z^2
                    +64576 z
                    -7872) \, ,
  \end{autobreak}\notag\\
  \begin{autobreak}
    \MoveEqLeft
    C^{P\,T}_7 = \frac{1}{1290240 \pi ^4 z (2 z-1)^5}
                    (245760 z^{13}
                    -778752 z^{12}
                    +612096 z^{11}
                    -280064 z^{10}
                    +2535808 z^9
                    -4643040 z^8
                    +1263696 z^7
                    +3706672 z^6
                    -2978528 z^5
                    -1258684 z^4
                    +2894850 z^3
                    -1666910 z^2
                    +437115 z
                    -44835) \, ,
  \end{autobreak}\notag\\
  \begin{autobreak}
    \MoveEqLeft
    C^{P\,T}_8 = -\frac{z}{80640 \pi ^4 (z-1)^5}
                    (480 z^{11}
                    -2721 z^{10}
                    +5598 z^9
                    -5137 z^8
                    +6864 z^7
                    -21665 z^6
                    +33666 z^5
                    -15377 z^4
                    -16968 z^3
                    +25816 z^2
                    -11620 z
                    +980) \, ,
  \end{autobreak}\notag\\
  \begin{autobreak}
    \MoveEqLeft
    C^{P\,T}_9 = -\frac{1}{10080 \pi ^4 z^2}
                    (1043 z^3
                    -1774 z^2
                    +1576 z
                    -492) \, ,
  \end{autobreak}\notag\\
  \begin{autobreak}
    \MoveEqLeft
    C^{P\,T}_{10} = -\frac{1}{5040 \pi ^4 z^2}
                      (1043 z^3
                      -1774 z^2
                      +1576 z
                      -492) \,
  \end{autobreak}\notag\\
  \begin{autobreak}
    \MoveEqLeft
  C^{P\,T}_{11} = \frac{1}{5040 \pi ^4 z^2}
                    (700 z^4
                    -1862 z^3
                    +2608 z^2
                    -1889 z
                    +488) \, .
  \end{autobreak}
\end{align}

\providecommand{\href}[2]{#2}\begingroup\raggedright\endgroup

%---------------------------------------------------------------------
%bibliography
%\bibliographystyle{utphysMa}
%\bibliography{bibTex1.4}
%\bibliography{E:/yqma/work/paper/tools/JabRef/bibTex1.4}
%\bibliography{C:/Work in C/gluon fragmentation/Fragmentation of 3S11/Paper/bibTex1.4}
%---------------------------------------------------------------------

\end{document}